\documentclass[1p,times,preprint]{elsarticle}
\usepackage[english]{babel}
\usepackage{tensor}
\usepackage{graphicx}
\usepackage{amsmath}
\usepackage{amssymb}
\usepackage{amsfonts}
\usepackage{dcolumn}
\usepackage{bm}
\usepackage{xcolor}
\usepackage{ulem}
\usepackage{tikz}
\usepackage{subcaption}
\usepackage{comment}
\usepackage{verbatim}
\usepackage{fancyvrb}
\usepackage{cancel}
\usepackage{multirow}
\usepackage{lscape}
\usepackage{txfonts}
\usepackage{mathtools}
\usepackage{soul}
\usepackage{url}
\usepackage{longtable}
\usepackage{makecell}
\usepackage[pdftex]{pict2e}
\usepackage{microtype}
\usepackage{float} 

\def\bra<#1|{\mathinner{\langle\,{#1}\,\vert}} 
\def\ket|#1>{\mathinner{\vert\,{#1}\,\rangle}} 
\def\red|#1|{\mathinner{\!\vert\,{#1}\,\vert\!}}
\def\braket<#1>{\mathinner{\langle\,{#1}\,\rangle}} 

\def\redmem#1#2#3{  \left\langle #1 \left\Vert  
                  #2 \right\Vert #3 \right\rangle   }

\begin{document}

\title{SECOND-ORDER RAYLEIGH-SCHR\"ODINGER PERTURBATION THEORY FOR THE {\sc Grasp}2018 PACKAGE: VALENCE-VALENCE CORRELATIONS}
\date{\today}

\author[TFAI]{G. Gaigalas}
\address[TFAI]{Institute of Theoretical Physics and Astronomy, 
               Vilnius University, Saul\.{e}tekio Ave. 3, LT-10257 Vilnius, Lithuania}
\ead{gediminas.gaigalas@tfai.vu.lt}

\author[TFAI]{P. Rynkun}
\ead{pavel.rynkun@tfai.vu.lt}

\author[TFAI]{L. Kitovien\.{e}}
\ead{laima.radziute@tfai.vu.lt}
 
%
%
\begin{abstract}
The accurate description of electron correlations remains a major challenge in atomic calculations. 
In order to perform accurate calculations, it is necessary to consider the various types 
of electron correlations what often leads to extensive CSF expansions.
This work presents further development of the method based on the second-order perturbation theory 
to identify the most significant CSFs that have the greatest influence on core-valence, core, core-core and valence-valence correlations. 
This method is based on a combination of the relativistic configuration interaction method 
 and the stationary second-order Rayleigh-Schr\"odinger many-body perturbation theory in an irreducible tensorial form
[G. Gaigalas, P. Rynkun, L. Kitovienė, 
Second-Order Rayleigh-Schr\"odinger Perturbation Theory for the {\sc Grasp}2018 Package: Core-Valence
Correlations, {\em Lithuanian Journal of Physics}, {\bf 64}, No. 1, 20-39 (2024) (https://doi.org/10.3952/physics.2024.64.1.3), 
G. Gaigalas, P. Rynkun, L. Kitovienė, 
Second-Order Rayleigh-Schr\"odinger Perturbation Theory for the {\sc Grasp}2018 Package: Core
Correlations, {\em Lithuanian Journal of Physics}, {\bf 64}, No. 2, 73-81 (2024) (https://doi.org/10.3952/physics.2024.64.2.1) 
and
G. Gaigalas, P. Rynkun, L. Kitovienė, 
Second-Order Rayleigh-Schr\"odinger Perturbation Theory for the {\sc Grasp}2018 Package: Core -Core
Correlations, {\em Lithuanian Journal of Physics}, {\bf 64}, No. 3, 139-161 (2024) 

\hspace{-0.55cm}(https://doi.org/10.3952/physics.2024.64.3.1)].
The method is extended to include additionally valence-valence electron correlations. 
It can be applied for an atom or ion with any number of valence electrons for calculation of energy spectra and other properties.
Meanwhile, the correlations which can not be included according to perturbation theory are accounted for in a regular way.
The use of the developed method
allows a significant reduction of CSFs especially for complex atoms and ions. 
As an example of its application, the atomic calculations of the energy structure for Se~III ion are presented.

\end{abstract}

\maketitle

\newpage

\section{Introduction}

Accurate calculation of atomic properties requires correlation and relativistic effects to be taken into account as much as possible. The
General Relativistic Atomic Structure ({\sc Grasp}) software package~\cite{grasp2013,grasp2018} is currently one of the most widely used software packages for such problems in atomic theory. It is well developed and has various options for capturing correlation effects as accurately as possible. But currently the demand for the accuracy of atomic quantities is very high, therefore the development of further methods and their implementation in software is needed. This series of papers \cite{Gaigetal:2024CV,Gaigetal:2024C,Gaigetal:2024CC} (the present paper included) aim to develop a method that builds on the conventional methods~\cite{Jonatal:2023a} already implemented in the {\sc Grasp} software package and, in addition, takes further advantage of perturbation theory (PT)~\cite{LindgrenBook:82,PT_book}.

Brillouin-Wigner perturbation theory (Zero-first-order method)~\cite{Jonatal:2023a,LindgrenBook:82,Gustatal:2017,Gaigtal:2020a,Jonatal:2023b} 
has been used (without energy iterative procedure) in the {\sc Grasp} software package for a number of years.
This latter version of the perturbation theory has a number of advantages~\cite[Subsection 1.3.3.1]{PT_book} over its other version, the Rayleigh-Schr\"odinger perturbation theory:
(i) Rayleigh–Schr\"odinger theory can be regarded as an approximation to Brillouin–Wigner theory,
(ii) Brillouin–Wigner theory is formally much simpler than the Rayleigh–Schr\"odinger theory,
(iii) Convergence of the Brillouin–Wigner perturbation theory is often more rapid than that of the Rayleigh–Schr\"odinger theory for a given problem,
(iv) Brillouin–Wigner perturbation theory may converge for problems for which the Rayleigh–Schr\"odinger theory does not,
(v) Brillouin–Wigner perturbation theory is often more convenient to use for degenerate problems.
It also has the disadvantages of the Brillouin-Wigner theory against Rayleigh-Schr\"odinger perturbation theory~\cite[Subsection 1.3.3.2]{PT_book}:
(i) Brillouin–Wigner perturbation theory is iterative, since the exact energy is contained in the denominators arising in the expressions for the energy components,
(ii) Brillouin–Wigner perturbation theory is not explicitly a many-body theory, in that the energy expressions in each order do not scale linearly with particle number.

The summary of the Rayleigh-Schr\"odinger many-body perturbation theory (RSMBPT) was given in the 1990s~\cite{LindgrenBook:82}.
It was then developed in the framework of non-relativistic atomic theory, using determinants.
Although this theory has some weaknesses, its strengths stimulate further development of the theory.
Its further modification resulted in irreducible tensorial form of effective Hamiltonian in the first two orders of PT~\cite{Meratal:85,Meratal:86,Gaigalas:89}.
The spin-angular part of Feynman diagrams in this version of theory is represented by the sum of irreducible tensorial products,
composed of two, four, and six creation and annihilation operators, which correspond to the Feynman diagrams with two, four, and six free lines~\cite{Gaigalas:89}. This allows to use Racah algebra~\cite{RI,Rudzikas_book:97} including quasispin~\cite{RudzKan_1984,Gaigalas_1996} for spin-angular integration in $LS$-coupling~\cite{Gaigalas_1997}.
But both the Rayleigh-Schr\"odinger perturbation theory~\cite{LindgrenBook:82} and the irreducible tensorial form of this theory~\cite{Meratal:85,Meratal:86,Gaigalas:89} are not suitable for use in the {\sc Grasp} software package because:
(a) {\sc Grasp} is based on relativistic atomic theory~\cite{grantBook:07,Jonatal:2023a} and Rayleigh-Schr\"odinger perturbation theory~\cite{LindgrenBook:82,Meratal:85,Meratal:86,Gaigalas:89} is based on non-relativistic atomic theory,
(b) the single electron orbitals in the {\sc Grasp} software package are searched for in an arbitrary potential (in multiconfiguration Dirac-Hartree-Fock approximation~\cite{grantBook:07}), which in the most general case does not satisfy the requirements of the Rayleigh-Schr\"odinger many-body perturbation theory~\cite{LindgrenBook:82}.
The latter problem (b) was successfully solved~\cite{Bogetal:1997,Bogetal:1998} by applying the theory in a non-relativistic framework of atomic theory using a superposition of configurations approach~\cite{fischer_brage}, where a \texttt{SAI} software library ~\cite{Gaigalas:2002} was used to search for the spin-angular coefficients in the $LS$-coupling~\cite{Gaigalas_1997}.
This series of articles~ \cite{Gaigetal:2024CV,Gaigetal:2024C,Gaigetal:2024CC} (including this one) is the final stage of development of this theory, i.e. both problems (a) and (b) are successfully solved at once.
This means that the RSMBPT theory is finally developed in such a way that it can easily be used in the {\sc Grasp} software package.

In fact the method developed in this paper is a combination of the relativistic configuration interaction (RCI) method and RSMBPT theory in the irreducible tensorial form (RCI (RSMBPT)), where core-valence (CV), core (C), core-core (CC), and valence-valence (VV) correlations are included according to the the stationary second-order Rayleigh-Schr\"odinger many-body perturbation theory.
It no longer suffers from disadvantage (ii) (see~\cite[Subsection 1.3.3.1]{PT_book}), since the theory is implemented in the {\sc Grasp} software package~\cite{grasp2013,grasp2018} and the user simply uses it as a "black box", inputting only the data of the system (atom or ion) under consideration.
The remaining drawbacks (i), (iii), (iv), and (v) (see~\cite[Subsection 1.3.3.1]{PT_book}) are also no longer relevant in the methodology proposed in this paper, as a number of correlation effects are included in the framework of multiconfiguration Dirac-Hartree-Fock approximation~\cite{grantBook:07,Fisetal:16a} already in the zero-order of perturbation theory.
Moreover, it is used in such a way that the methodology is not used for the final calculations, but rather for the assessment of the influence of correlations -- which correlations have the largest contribution to the final results, and which correlations need to be included in the final RCI calculations.
In addition, the methodology proposed in this work, RCI (RSMBPT), also without restrictions, allows the use of the Brillouin-Wigner theory (Zero-first-order method  without energy iterative procedure)~\cite{Jonatal:2023a,LindgrenBook:82,Gustatal:2017,Gaigtal:2020a,Jonatal:2023b}, which is already implemented in the {\sc Grasp} software package~\cite{grasp2013,grasp2018}.
So now, of the (i)-(v) disadvantages of the Rayleigh-Schr\"odinger perturbation theory listed in \cite[Subsection 1.3.3.1]{PT_book}, only (i) and (ii) advantages from~\cite[Subsection 1.3.3.2]{PT_book} remain.

All these expressions for the Feynman diagrams corresponding
to valence-valence correlations in relativistic atomic theory are first presented in Section~\ref{sec:seconOrder}.
The modified expressions of the Feynman diagrams which are
able to use the spin-angular program library \texttt{librang}~\cite{Gaigalas:2022} from {\sc Grasp}~\cite{grasp2013,grasp2018} with any further modifications
are given in Section~\ref{Sec:VVImplementation}.
The validity and efficiency of the presented RCI (RSMBPT) method is demonstrated in Section~\ref{Sec:Calculations}, where the energy
spectrum of the Se~III ion are theoretically studied.

\section{Relativistic second order effective Hamiltonian of an atom or ion in irreducible tensorial form for valence-valence correlations}
\label{sec:seconOrder}

New Feynman diagrams are used to describe valence-valence correlations, which were not available for core-valence, core, and core-core correlations. We discuss this in detail below.

\subsection{The first type of valence-valence correlations}
\label{subsec:VVfirst_type}

This type of valence-valence correlations is presented through the Feynman diagrams VV$_1$ from Fig.~\ref{VV_1} and VV$_2$ from Fig.~\ref{VV_2} where all lines with double arrow of diagrams are renamed $m$ i.e. $m'=n=n' \equiv m$.
\begin{equation}
\label{eq:VV-a} 
(n_{m} \ell_{m}) \; j_{m}^{w_m} \; (n_{n} \ell_{n}) \; j_{n}^{w_n}
   \rightarrow (n_{m} \ell_{m})\, j_{m}^{w_m-2} \; (n_{n} \ell_{n}) \; j_{n}^{w_n}
	 \; (n_{r} \ell_{r}) \; j_{r} \; (n_{s} \ell_{s}) \; j_{s} ,
\end{equation}
\begin{equation}
\label{eq:VV-a2} 
(n_{m} \ell_{m}) \; j_{m}^{w_m} \; (n_{n} \ell_{n}) \; j_{n}^{w_n}
   \rightarrow (n_{m} \ell_{m})\, j_{m}^{w_m-2} \; (n_{n} \ell_{n}) \; j_{n}^{w_n}
	 \; (n_{s} \ell_{s})\, j_{s}^2 .
\end{equation}

\begin{figure*}
\begin{center}
\setlength{\unitlength}{1mm}
\begin{picture}(180,43)
\thicklines
\put(10,30){\line(0,1){10}}
\put(10,38){\vector(0,1){2}}
\put(10,40){\vector(0,1){2}}
\put(12.5,38){\makebox(0,0)[t]{\small{$m$}}}
\put(20,30){\line(-1,-1){10}}
\put(18,22){\vector(-1,1){2}}
\put(11,27){\makebox(0,0)[t]{\small{$r$}}}
\multiput(10,30)(1,0){10}{\circle*{0.35}}
\put(20,20){\line(0,10){10}}
\put(20,26.5){\vector(0,-1){2}}
\put(20,24.5){\vector(0,-1){2}}
\put(22.5,25){\makebox(0,0)[t]{\small{$n$}}}
\multiput(10,20)(1,0){10}{\circle*{0.35}}
\put(10,10){\line(0,1){10}}
\put(10,10){\vector(0,1){3}}
\put(12.5,13){\makebox(0,0){\small{$m^{\prime}$}}}
\put(10,08){\vector(0,1){3}}
\put(20,20){\line(-1,1){10}}
\put(16,26){\vector(1,1){2}}
\put(11,24){\makebox(0,0)[t]{\small{$s$}}}
\put(15,04){\makebox(0,0){$\text{VV}_{1}$}}
\put(27,25){\makebox(0,0) [l] {$\displaystyle{ = -
\frac{1}{2} \; \sum_{m, m^{\prime}, n}~\frac{1}{\sqrt{\left[ j_{m} \right]}}~\left[\;  a^{\left( j_m \right) } \times
  \tilde a^{\left( j_{m^{\prime}} \right) } \; \right] ^{\left( 0 \right)}
  ~\sum_{r , s}~\frac{\left( -1 \right)^{j_m+j_n+j_{r}+j_s}}{\left( \varepsilon_{m'}+\varepsilon_n-\varepsilon_r-\varepsilon_s \right)}
}$}}
\put(37,10){\makebox(0,0) [l] {$\displaystyle{ \times
  \sum_{k,k'}
  \left\{
    \begin{array}{ccc}
      k  & j_{m} & j_{r} \\
      k' & j_{n} & j_{s}
    \end{array} \right\} 
	~X_{k}(m n, r s) ~ X_{k'}(s r, m^{\prime} n)}$}}
\end{picture}
\caption{The VV one-particle Feynman diagram of the second-order effective Hamiltonian for valence-valence correlations.}
\label{VV_1}
\end{center}
\end{figure*}
\begin{figure}
\begin{center}
\setlength{\unitlength}{1mm}
\begin{picture}(180,58)
\thicklines
\put(10,45){\vector(0,1){10}}
\put(10,55){\vector(0,1){2}}
\put(7.5,53){\makebox(0,0)[t]{\small{$m$}}}
\put(20,45){\vector(0,1){10}}
\put(20,55){\vector(0,1){2}}
\put(22,53){\makebox(0,0)[t]{\small{$n$}}}
\multiput(10,45)(1,0){10}{\circle*{0.35}}
\put(10,35){\line(0,10){10}}
\put(10,39){\vector(0,1){2}}
\put(7.5,43){\makebox(0,0)[t]{\small{$r$}}}
\put(20,35){\line(0,10){10}}
\put(20,39){\vector(0,1){2}}
\put(22,43){\makebox(0,0)[t]{\small{$s$}}}
\multiput(10,35)(1,0){10}{\circle*{0.35}}
\put(10,25){\line(0,1){10}}
\put(10,25){\vector(0,1){3}}
\put(7.5,28){\makebox(0,0){\small{$m'$}}}
\put(10,23){\vector(0,1){3}}
\put(20,25){\line(0,1){10}}
\put(20,25){\vector(0,1){3}}
\put(15,19){\makebox(0,0){$ VV_{2}$}}
\put(22,28){\makebox(0,0){\small{$n'$}}}
\put(20,23){\vector(0,1){3}}
\put(26,40){\makebox(0,0) [l] {$\displaystyle{ =
\frac{1}{2} \; \sum_{m, m^{\prime}}~\sum_{n, n^{\prime}}~\sum_{j_{12}}~~\sqrt{\left[ j_{12} \right]}~ 
		\left[\left[\;  a^{\left( j_m \right) }  \times 
  \tilde a^{\left( j_{m^{\prime}} \right) } \; \right] ^{\left( j_{12} \right)} \times 
	\left[\; a^{\left( j_{n} \right)}  \times \tilde a^{\left( j_{n^{\prime}} \right) }
  \; \right] ^{\left( j_{12} \right)} \right]^{\left( 0 \right)}	
	}$}}
\put(36,25){\makebox(0,0) [l] {$\displaystyle{ \times
\sum_{r, s}~\sum_{k,k'}~\frac{\left( -1 \right)^{j_{m}+j_{m^{\prime}}+j_{n}+j_{n^{\prime}}}}
{\left( \varepsilon_{m'}+\varepsilon_{n'}-\varepsilon_r-\varepsilon_s \right)}
	\left\{
    \begin{array}{ccc}
      k  & k' & j_{12} \\
      j_{m'} & j_{m} & j_{r}
     \end{array}
	\right\} 
  \left\{
    \begin{array}{ccc}
      k  & k' & j_{12} \\
      j_{n'} & j_{n} & j_{s}
    \end{array}
	\right\}
	}$}}
\put(36,10){\makebox(0,0) [l] {$\times \; X_{k}(m n, r s) ~ X_{k'}(r s, m^{\prime} n^{\prime})$}}
\end{picture}
\caption{The VV two-particle Feynman diagram of the second-order effective Hamiltonian for valence-valence correlations.}
\label{VV_2}
\end{center}
\end{figure}

Each second order Feynman diagram's expression of Rayleigh-Schr\"odinger  many-body perturbation theory as mentioned in~\cite{Gaigetal:2024CV,Gaigetal:2024C,Gaigetal:2024CC}
 has the energy denominator
$D = \sum \left( \varepsilon_{\text{down}} - \varepsilon_{\text{up}} \right)$,
where $\varepsilon_{\text{down}}$ ($\varepsilon_{\text{up}}$) is the single-particle eigenvalue associated with the down- (up-)
orbital lines to (from) the lowest interaction line of the diagram. For example the denominators for VV$_1$ diagram are
\begin{equation}
\label{eq:denominator}
D = \left( \varepsilon_{m'}+\varepsilon_n-\varepsilon_r-\varepsilon_s \right) ,
\end{equation}
where indexes $m'$ and $n$ belong to $F'$ set, and $r$, $s$ belong to $G$ set of orbitals~\cite{Gaigetal:2024CV}. 

Also, the following notations are used in the expressions of these diagrams (see Fig.~\ref{VV_1} and Fig.~\ref{VV_2}):
\begin{equation}
\label{eq:deffX}
   X_{k}(i j, i' j') 
   = \redmem{\ell_i j_{i}}{\, C^{(k)} \,}{ \ell_{i'} j_{i'}}
     \redmem{\ell_j j_{j}}{\, C^{(k)} \,}{ \ell_{j'} j_{j'}} 
R^{k}(n_i j_i \, n_jj_j, \, n_{i'}j_{i'} \, n_{j'}j_{j'} ) ,
\end{equation}
where $R^{k}\left(n_i j_i \, n_jj_j, \, n_{i'}j_{i'} \, n_{j'}j_{j'} \right)$ is the radial integral 
of electrostatic interaction between electrons~\cite[(89) and (90)]{Fisetal:16a} and 
$\redmem{\ell_i j_{i}}{\, C^{(k)} \,}{ \ell_{i'} j_{i'}}$ is the reduced matrix element of the irreducible tensor operator $C^{(k)}$ in $jj$-coupling.

Depending on how the Feynman diagrams VV$_1$ and VV$_2$ act on the open subshells (orbitals from $F'$ set), the operators of second quantization (open lines with double arrow in diagrams) take on corresponding values, i.e. specific values for the summation parameters $m$, $m'$, $n$, and $n'$. In addition, in the diagram VV$_1$, the summation parameter $n$ passes through all possible orbitals that belong to the $F'$ set.
All these specific values of the parameters $m$, $m'$, $n$, and $n'$ determine 
the type of valence-valence correlations represented by the Feynman diagrams.
We also note that in the algebraic expressions of the both diagrams VV$_1$ and VV$_2$, the sum over excited orbits
$n_r \ell_{r} j_{r}$ and $n_s \ell_{s} j_{s}$ runs.

The diagram VV$_1$ is a one-particle Feynman diagram where its tensorial part is expressed through tensorial product of creation $a^{\left( j \right) }$ and annihilation $\tilde a^{\left( j\right) }$ operators
\begin{equation}
\label{eq:s-a_VV3}
 \left[\;     a^{\left( j_m \right) }  \times
\tilde a^{\left( j_{m^{\prime}} \right) } \; \right] ^{\left( 0 \right)} .
\end{equation}
These two operators of second quantization are in normal form, act on the same subshell $m$, and represent the scalar operator
for this first type (\ref{eq:VV-a}) and (\ref{eq:VV-a2}) of valence-valence correlations. Therefore, this operator can be expressed through the operator of subshell occupation number $\hat{N}$ on which it acts
\begin{equation}
\label{eq:s-b_VV3}
 \left[\; a^{\left( j_m \right) }  \times
   \tilde a^{\left( j_{m}  \right) } \; \right] ^{\left( 0 \right)} = -\, \frac{\hat{N}_m}{\sqrt{\left[ j_m \right]}} .
\end{equation}
There is no need to use the spin-angular program library \texttt{librang}~\cite{Gaigalas:2022} to calculate this diagram, as the spin-angular coefficient is expressed through a simple multiplier~~\cite[A3]{Gaigetal:2005}. This is one of the advantages of the methodology proposed in this paper.

The VV$_2$ is a two-particle Feynman diagram where its tensorial part is expressed through tensorial product of creation $a^{\left( j \right) }$ and annihilation $\tilde a^{\left( j\right) }$ operators and represent the scalar operator (see ~\cite[Section 2.3 and Fig. 2]{Gaigetal:2024CV} for more details)
\begin{eqnarray}
\label{eq:VV3-c}
	\left[\left[\;  a^{\left( j_m \right) } \times 
   \tilde a^{\left( j_{m^{\prime}} \right) } \right] ^{\left( j_{12} \right)} \times 
	\left[\; a^{\left( j_n \right) }  \times 
  \tilde a^{\left( j_{n^{\prime}} \right) }  \right] ^{\left( j_{12} \right)} \right]^{\left( 0 \right)} .
\end{eqnarray}
For this VV$_2$ diagram, unlike for the VV$_1$ Feynman diagram considered above, there comes the additive summation, where the summation parameter $j_{12}$ relates the tensor product (\ref{eq:VV3-c}) to the two $6j$- coefficients. This summation parameter is the intermediate rank of this tensor product.
All operators of second quantization act on the same subshell $m$ therefore the tensorial operator
(\ref{eq:VV3-c}) can be expressed through simple two $\hat{N}$ operators in case $j_{12}=0$
\begin{eqnarray}
\label{eq:VV3-d}
	\left[\left[\;  a^{\left( j_m \right) } \times 
  \tilde a^{\left( j_m \right) } \; \right] ^{\left( 0 \right)} \times 
	\left[\; a^{\left( j_m \right) }  \times 
  \tilde a^{\left( j_m \right) } \; \right] ^{\left( 0 \right)} \right]^{\left( 0 \right)}
	= \frac{\hat{N}_m^2}{\left[ j_m \right]} .
\end{eqnarray}
So for this case, for calculation of VV$_2$ diagram, we also do not
need to use the spin-angular program library \texttt{librang}~\cite{Gaigalas:2022}. Meanwhile,
the program library \texttt{librang}~\cite{Gaigalas:2022} from {\sc Grasp} supports calculation of spin-angular part (\ref{eq:VV3-c}) of VV$_2$ Feynman diagram in case $j_{12}>0$. Full Racah algebra~\cite{RI} including quasispin~\cite{Gaigalas_1997} is available for integration of spin-angular part of this Feynman diagram (see \cite[(29) and (30)]{Gaigalas:2022}).
This is another advantage of the methodology proposed in this paper.

\subsection{The second type of valence-valence correlations}
\label{subsec:VVsecond_type}

This type of valence-valence correlations is presented through the Feynman diagram VV$_2$ from Fig.~\ref{VV_2} where all lines with double arrow of diagrams are renamed in the following way: $m'\equiv m$ and $n' \equiv n$.
\begin{equation}
\label{eq:VV-b} 
(n_{m} \ell_{m}) \; j_{m}^{w_m} \; (n_{n} \ell_{n}) \; j_{n}^{w_n}
   \rightarrow (n_{m} \ell_{m})\, j_{m}^{w_m-1} \; (n_{n} \ell_{n}) \; j_{n}^{w_n-1}
	 \; (n_{r} \ell_{r}) \; j_{r} \; (n_{s} \ell_{s}) \; j_{s} ,
\end{equation}
\begin{equation}
\label{eq:VV-b2} 
(n_{m} \ell_{m}) \; j_{m}^{w_m} \; (n_{n} \ell_{n}) \; j_{n}^{w_n}
   \rightarrow (n_{m} \ell_{m})\, j_{m}^{w_m-1} \; (n_{n} \ell_{n}) \; j_{n}^{w_n-1}
	 \; (n_{s} \ell_{s})\, j_{s}^2 .
\end{equation}

In this case the tensorial part of the VV$_2$ diagram has the following form:
\begin{eqnarray}
\label{eq:VV4-c}
	\left[\left[\;  a^{\left( j_m \right) } \times 
  \tilde a^{\left( j_m \right) } \; \right] ^{\left( j_{12} \right)} \times 
	\left[\;  a^{\left( j_n \right) } \times 
  \tilde a^{\left( j_n \right) } \; \right] ^{\left( j_{12} \right)} \right]^{\left( 0 \right)} ,
\end{eqnarray}
which can be expressed in case $j_{12} = 0$  as
\begin{eqnarray}
\label{eq:VV4-d}
	\left[\left[\;  a^{\left( j_m \right) } \times 
  \tilde a^{\left( j_m \right) } \; \right] ^{\left( 0 \right)} \times 
	\left[\;  a^{\left( j_n \right) } \times 
  \tilde a^{\left( j_n \right) } \; \right] ^{\left( 0 \right)} \right]^{\left( 0 \right)}
	= \frac{\hat{N}_m  \hat{N}_n}{\sqrt{\left[ j_m, j_n \right]}} .
\end{eqnarray}
So for this case we also do not
need to use the spin-angular program library \texttt{librang}~\cite{Gaigalas:2022} for calculation of the VV$_2$ diagram. Meanwhile,
the program library \texttt{librang}~\cite{Gaigalas:2022} from {\sc Grasp} supports calculation of the spin-angular part (\ref{eq:VV4-c}) of the VV$_2$ Feynman diagram in case $j_{12}>0$. Full Racah algebra~\cite{RI} including quasispin~\cite{Gaigalas_1997} is available for integration of the spin-angular part of this Feynman diagram (see \cite[(26), (12), and (13)]{Gaigalas:2022}).

\subsection{The contribution of valence-valence correlations to off-diagonal matrix elements}
\label{subsec:offdiagonal_type}

The main contribution of valence-valence correlations to off-diagonal matrix elements is in the matrix element
$\redmem{(n_{m} \ell_{m})\, j_{m}^{w_m} \, (n_{n} \ell_{n})\, j_{n}^{w_n}}{\, \widehat{{\cal H}}^{(2)}_{\text{Effective}} \,}{(n_{m} \ell_{m})\, j_{m}^{w_m-2} \, (n_{n} \ell_{n})\, j_{n}^{w_n+2} }$.
The above contribution is derived from the excitation:
\begin{equation}
\label{eq:VV-off_Diagonal_a} 
(n_{m} \ell_{m}) \; j_{m}^{w_m} \; (n_{n} \ell_{n}) \; j_{n}^{w_n}
   \rightarrow (n_{m} \ell_{m})\, j_{m}^{w_m-2} \; (n_{n} \ell_{n}) \; j_{n}^{w_n}
	 \; (n_{r} \ell_{r}) \; j_{r} \; (n_{s} \ell_{s}) \; j_{s}
\end{equation}
and can be described through the same two-particle Feynman diagram from Fig.~\ref{VV_2}, as above. 
This type of valence-valence correlations is presented through the Feynman diagram VV$_2$
where all lines with double arrow of diagram are renamed in the following way: $m'\equiv n$, $n \equiv m$ and $n' \equiv n$.

In this case, the tensorial part of the VV$_2$ diagram has the following form:
\begin{eqnarray}
\label{eq:VV-off_Diagonal_b}
	\left[\left[\;  a^{\left( j_n \right) } \times 
  \tilde a^{\left( j_m \right) } \; \right] ^{\left( j_{12} \right)} \times 
	\left[\;  a^{\left( j_n \right) } \times 
  \tilde a^{\left( j_m \right) } \; \right] ^{\left( j_{12} \right)} \right]^{\left( 0 \right)} .
\end{eqnarray}
In this case, after some modifications similar to~\cite{Gaigalas_1996} made in Section~\ref{Sec:VVImplementation}, the Racah algebra~\cite{RI,Gaigalas_1997} and the software
library \texttt{librang}~\cite{Gaigalas:2022} are also fully available (see \cite[(26), (12), and (13)]{Gaigalas:2022}).

\subsection{The remaining type of valence-valence correlations}
\label{subsec:remaining_type}

The remaining types of valence-valence correlations
\begin{equation}
\label{eq:remaining_VV-a}
    (n_{m} \ell_{m})\, j_{m}^{w_m} \; (n_{n} \ell_{n})\, j_{n}^{w_n} 
   \rightarrow (n_{m} \ell_{m})\, j_{m}^{w_m+1} \; (n_{n} \ell_{n})\, j_{n}^{w_n-2} \; (n_{r} \ell_{r})\, j_{r} , 
\end{equation}
\begin{equation}
\label{eq:remaining_VV-b}
    (n_{m} \ell_{m})\, j_{m}^{w_m} \; (n_{n} \ell_{n})\, j_{n}^{w_n} \; (n_{p} \ell_{p})\, j_{p}^{w_p} 
   \rightarrow (n_{m} \ell_{m})\, j_{m}^{w_m+1} \; (n_{n} \ell_{n})\, j_{n}^{w_n-1}  \; (n_{p} \ell_{p})\, j_{p}^{w_p-1} \; (n_{r} \ell_{r})\, j_{r} 
\end{equation}
are impossible to be included via RSMBPT without improvement of the spin-angular program library \texttt{librang}~\cite{Gaigalas:2022}. This is related to the fact that these types of correlations are described by three-particle Feynman diagram. The index $p$ in the expression (\ref{eq:remaining_VV-b}) belongs to space $F'$, as do the indices $n$ and $m$.

\section{Combination of RCI approximation
with the stationary second-order Rayleigh-Schr\"odinger many-body perturbation theory}
\label{Sec:VVImplementation}

Similar to CV, C, and CC correlations \cite{Gaigetal:2024CV,Gaigetal:2024C,Gaigetal:2024CC}, the admixed configurations from VV correlations 
(Eqs. (\ref{eq:VV-a}), (\ref{eq:VV-a2}), (\ref{eq:VV-b}), and (\ref{eq:VV-b2}))
can be added to usual energy $E_0 \left(K \right)$ of the 
term $\chi J$ of the configuration $K$ and can be
expressed as the energy $\Delta \mathcal{E}_0 \left(K J \right)$, which does not depend on the term, and the sum of the product of Slater integrals and spin-angular coefficients, describing the interaction within open subshells and between them:
\begin{eqnarray}
\label{eq:VVBogEnergy}
\hspace*{-2.5cm}
   E\left(K \chi J \right)
	\nonumber \\
& &
   = E_0 \left(K J\right) + \Delta \mathcal{E}_0 \left(K J \right) 
	\nonumber \\  [0.2cm]
& &
	+ \; \sum_{n\ell j} \sum_{k>0} \widetilde{f}_k \left( \ell j^{w}, \; K \chi J  \right)
	\left[ \mathcal{F}^{k} \left( n \ell j, \; n \ell j \right)  
	+ \Delta \mathcal{F}^{k} \left( n \ell j, \; n \ell j \right) \right]
	\nonumber \\
& &
	+ \; \sum_{n\ell j} \sum_{n'\ell'j' > n\ell j} \left\{ \sum_{k>0} \widetilde{f}_k \left( \ell j^{w} \; \ell' j'^{w'},
	\; K \chi J  \right) \right.
\left[ \mathcal{F}^{k} \left( n \ell j, \; n '\ell' j' \right)  
	+ \Delta \mathcal{F}^{k} \left( n \ell j, \; n' \ell' j' \right) \right]
	\nonumber \\ [0.2cm]
& & 
	+ \sum_{k} \widetilde{g}_k \left( \ell j^{w} \; \ell' j'^{w'}, \; K \chi J  \right)	
\left[ \mathcal{G}^{k} \left( n \ell j, \; n '\ell' j' \right)  
	+ \Delta \mathcal{G}^{k} \left( n \ell j, \; n' \ell' j' \right) \right] 
	\nonumber \\ [0.2cm]
& &
	+ \sum_{k} \widetilde{v}_k \left( \ell j^{w} \; \ell' j'^{w'}, \ell j^{w-2} \; \ell' j'^{w'+2},
	\; K \chi J \; K' \chi' J \right)	
	\nonumber \\
& &
  \times 
\left. 
\left[ \mathcal{R}^{k} \left( n \ell j n \ell j, \; n '\ell' j' n '\ell' j' \right)  
	+ \Delta \mathcal{R}^{k} \left( n \ell j n \ell j, \; n' \ell' j' n '\ell' j' \right) \right] \right\} ,
\end{eqnarray}
where $\widetilde{f}_k$, $\widetilde{g}_k$, and $\widetilde{v}_k$ are spin-angular coefficients from which submatrix elements $\redmem{\ell j}{\, C^{(k)} \,}{ \ell^{\prime} j^{\prime}}$ are extracted. 
Therefore summation over $k$ runs over all 
possible values instead of the values which satisfy the triangular condition $\left( \ell \ell^{\prime} k\right)$ as it is in the ordinary case. The $\mathcal{F}^{k} \left( n \ell j, \; n '\ell' j' \right)$,
$\mathcal{G}^{k} \left( n \ell j, \; n '\ell' j' \right)$, and $\mathcal{R}^{k} \left( n \ell j n \ell j, \; n '\ell' j' n '\ell' j' \right)$ are generalized integrals of electrostatic interaction between electrons. The definition of $\mathcal{R}^{k} \left( n \ell j n \ell j, \; n '\ell' j' n '\ell' j' \right)$ is the following:
\begin{eqnarray}
\label{eq:BogRk}
\hspace*{-2.5cm}
   \mathcal{R}^{k}\left(i j, i' j'\right)
	\nonumber \\
& &
   = \left\{ \left[ 1 + \delta \left( i, j \right) \right]  \left[ 1 + \delta \left( i', j' \right) \right]  \right\} ^{-1/2}
	 \, R^{k}\left(n_i j_i \, n_jj_j, \, n_{i'}j_{i'} \, n_{j'}j_{j'} \right)
\nonumber \\
& &
	   \times \redmem{\ell_i j_{i}}{\, C^{(k)} \,}{ \ell_{i'} j_{i'}}
     \redmem{\ell_j j_{j}}{\, C^{(k)} \,}{ \ell_{j'} j_{j'}}, 
\end{eqnarray}
where $R^{k}\left(n_i j_i \, n_jj_j, \, n_{i'}j_{i'} \, n_{j'}j_{j'} \right)$ is the same radial integral as in 
Eq. (\ref{eq:deffX}). Definitions  $\mathcal{F}^{k} \left( n \ell j, \; n '\ell' j' \right)$,
$\mathcal{G}^{k} \left( n \ell j, \; n '\ell' j' \right)$ straightforwardly follow from Eq. (\ref{eq:BogRk}).

The contribution coming from the VV correlations of the configurations $K'$ to $E (K \chi J)$ in the second order of the perturbation theory
can be written from Eq. (\ref{eq:VVBogEnergy}) as
\begin{eqnarray}
\label{eq:BogEnergy_PT}
\hspace*{-2.5cm}
  \Delta E_{PT (VV)}
	\nonumber \\
& &
   =  \Delta \mathcal{E}_0 \left(K J \right) 
	\nonumber \\  [0.2cm]
& &
	+ \; \sum_{n\ell j} \sum_{k>0} \widetilde{f}_k \left( \ell j^{w}, \; K \chi J  \right)
	\Delta \mathcal{F}^{k} \left( n \ell j, \; n \ell j \right)
	\nonumber \\
& &
	+ \; \sum_{n\ell j} \sum_{n'\ell'j' > n\ell j} \left\{ \sum_{k>0} \widetilde{f}_k \left( \ell j^{w} \; \ell' j'^{w'},
	\; K \chi J  \right) \right.
 \Delta \mathcal{F}^{k} \left( n \ell j, \; n' \ell' j' \right) 
	\nonumber \\ [0.2cm]
& & 
	+ \sum_{k} \widetilde{g}_k \left( \ell j^{w} \; \ell' j'^{w'}, \; K \chi J  \right)	
 \Delta \mathcal{G}^{k} \left( n \ell j, \; n' \ell' j' \right)
	\nonumber \\ [0.2cm]
& &
\left.
	+ \sum_{k} \widetilde{v}_k \left( \ell j^{w} \; \ell' j'^{w'}, \ell j^{w-2} \; \ell' j'^{w'+2},
	\; K \chi J \; K' \chi' J \right)	
\Delta \mathcal{R}^{k} \left( n \ell j n \ell j, \; n' \ell' j' n '\ell' j' \right) \right\} .
\end{eqnarray}

\begin{table*}
\begin{center}
\begin{tabular}{|l|} \hline
$\Delta \mathcal{E}_0$ corrections \\ \hline \hline
\\
$\overbrace{(n_{m} \ell_{m})\, j_{m}^{w_m} \; (n_{n} \ell_{n})\, j_{n}^{w_n}}^{\text{valence subshells}} \;
\rightarrow \; \overbrace{(n_{m} \ell_{m})\, j_{m}^{w_m-2} \; (n_{n} \ell_{n})\, j_{n}^{w_n}}^{\text{valence subshells}} \; \overbrace{(n_{r} \ell_{r})\, j_{r} \; (n_{s} \ell_{s})\, j_{s}}^{\text{virtual subshells}}$ \\
\\
$\underbrace{- \frac{2 w_m \left( w_m - 1 \right)}{\left[ j_m\right]} \; \mathcal{A'}\left( 0, \, m m, \, r s \right)
	}_{\text{from $\text{VV}_1$ and $\text{VV}_2$ Feynman diagrams}}$ \\
\\ \hline \hline
\\
$\overbrace{(n_{m} \ell_{m})\, j_{m}^{w_m} \; (n_{n} \ell_{n})\, j_{n}^{w_n}}^{\text{valence subshells}} \;
\rightarrow \; \overbrace{(n_{m} \ell_{m})\, j_{m}^{w_m-1} \; (n_{n} \ell_{n})\, j_{n}^{w_n-1}}^{\text{valence subshells}} \; \overbrace{(n_{r} \ell_{r})\, j_{r} \; (n_{s} \ell_{s})\, j_{s}}^{\text{virtual subshells}}$  \\
\\
$\underbrace{- \frac{w_m w_n}{\sqrt{\left[ j_m, j_n \right]}} \; \left[ \mathcal{A'}\left( 0, \, m n, \, r s \right) +\mathcal{A'}\left( 0, \, m n, \, s r \right) \right]}_{\text{from $\text{VV}_2$ Feynman diagram}}$ \\
\\ \hline 
\end{tabular}
\end{center}
\caption{Expressions for both two types (where $s \neq r$ or $s = r$) of valence-valence corrections to the energy in Eq. (\ref{eq:VVBogEnergy}), independent of the term.}
\label{tab:Implemen_VV1}
\end{table*}

\begin{table*}
\begin{center}
\begin{tabular}{|lll|} \hline
Corrections & Slater integral & $k$ values\\ \hline  \hline
& & \\
\multicolumn{3}{|c|}{$\overbrace{(n_{m} \ell_{m})\, j_{m}^{w_m} \; (n_{n} \ell_{n})\, j_{n}^{w_n}}^{\text{valence subshells}} \;
\rightarrow \; \overbrace{(n_{m} \ell_{m})\, j_{m}^{w_m-2} \; (n_{n} \ell_{n})\, j_{n}^{w_n}}^{\text{valence subshells}} \; \overbrace{(n_{r} \ell_{r})\, j_{r} \; (n_{s} \ell_{s})\, j_{s}}^{\text{virtual subshells}}$} \\
& &\\
$\underbrace{ 
-4 \; \left[ k \right] \mathcal{A^{\prime}}\left( k, \, m m, \, r s \right) }_{\text{from $\text{VV}_{1}$ and $\text{VV}_{2}$ Feynman diagrams}}$ \hspace{2cm} 
&$\Delta \mathcal{F}^{k}(m,m)$ & $k>0$\\
& &\\ \hline \hline
 & &\\
\multicolumn{3}{|c|}{$\overbrace{(n_{m} \ell_{m})\, j_{m}^{w_m} \; (n_{n} \ell_{n})\, j_{n}^{w_n}}^{\text{valence subshells}} \;
\rightarrow \; \overbrace{(n_{m} \ell_{m})\, j_{m}^{w_m-1} \; (n_{n} \ell_{n})\, j_{n}^{w_n-1}}^{\text{valence subshells}} \; \overbrace{(n_{r} \ell_{r})\, j_{r} \; (n_{s} \ell_{s})\, j_{s}}^{\text{virtual subshells}}$} \\
& &\\
$\underbrace{ 
- \; \left[ k \right] \left( \mathcal{A^{\prime}}\left( k, \, m n, \, r s \right) + \mathcal{A^{\prime}}\left( k, \, m n, \, s r \right) \right)}_{\text{from $\text{VV}_{2}$ Feynman diagram}}$ 
&$\Delta \mathcal{F}^{k}(m,n)$ & $k>0$\\
& &\\
& &\\
$\underbrace{ 
- 2\; \left[ k \right] \mathcal{B}\left( k, \, m n, \, r s \right)}_{\text{from $\text{VV}_{2}$ Feynman diagram}}$ 
&$\Delta \mathcal{G}^{k}(m,n)$ & $k\geq 0$\\
& & \\ \hline 
\end{tabular}
\end{center}
\caption{Expressions for Slater integrals $\Delta \mathcal{F}^{k}(m,m)$, $\Delta \mathcal{F}^{k}(m,n)$, and $\Delta \mathcal{G}^{k}(m,n)$ 
(see Eq. (\ref{eq:VVBogEnergy})) corresponding 
to both two types (where $s \neq r$ or $s = r$) of valence-valence corrections.} 
\label{tab:Implemen_VV2}
\end{table*}

The contribution of the first and second type of the VV correlations in the second-order of the perturbation theory is expressed over $\Delta \mathcal{E}_0 \left(K J \right)$, $\Delta \mathcal{F}^{k} \left( n \ell j, \; n \ell j \right)$, $\Delta \mathcal{F}^{k} \left( n \ell j, \; n' \ell' j' \right)$,
and $\Delta \mathcal{G}^{k} \left( n \ell j, \; n' \ell' j' \right)$ (see 
Tables~\ref{tab:Implemen_VV1} and \ref{tab:Implemen_VV2}).
These formulae are additionally expressed via the quantities

\begin{equation}
\label{eq:BogAp}
   \mathcal{A^{\prime}}\left(x, \; i j, \; i' j'\right) 
  = \sum_{k,k'}
		  \left\{
    \begin{array}{ccc}
      k  & k' & x \\
      j_{i} & j_{i} & j_{i'}
    \end{array} \right\}
				  \left\{
    \begin{array}{ccc}
      k  & k' & x \\
      j_{j} & j_{j} & j_{j'}
    \end{array} \right\}
\mathcal{P}\left(kk', \; i j, \; i' j'\right) ,
\end{equation}
and
\begin{equation}
\label{eq:BogB}
   \mathcal{B}\left(x, \; i j, \; i' j'\right) 
  = \sum_{k,k'}
		  \left\{
    \begin{array}{ccc}
      k  & k' & x \\
      j_{j} & j_{i} & j_{i'}
    \end{array} \right\}
				  \left\{
    \begin{array}{ccc}
      k  & k' & x \\
      j_{i} & j_{j} & j_{j'}
    \end{array} \right\}
\mathcal{Q}\left(kk', \; i j, \; i' j'\right) ,
\end{equation}
where
\begin{equation}
\label{eq:BogP}
   \mathcal{P}\left(kk', \; i j, \; i' j'\right) 
   = \mathcal{R}^{k}\left(i j, \; i' j'\right) \; \mathcal{R}^{k'}\left(i' j', \; i j \right)  \; 
	\mathcal{O}\left(K', K \right) ,
\end{equation}

\begin{equation}
\label{eq:BogQ}
   \mathcal{Q}\left(kk', \; i j, \; i' j'\right)
   = \mathcal{R}^{k}\left(i j, \; i' j'\right) \; \mathcal{R}^{k'}\left(i' j', \; j i\right)  \; 
\mathcal{O}\left(K', K \right) ,
\end{equation}

\begin{equation}
\label{eq:BogO1}
\mathcal{O}\left(K', K \right)
= \frac{1}{\overline{E}\left(K' \right)-\overline{E}\left(K\right)} ,
\end{equation}
where $\overline{E}\left(K\right)$ is the averaged energy of the 
state 
for which calculations are performed. $\overline{E}\left(K^{'}\right)$ is the averaged energy for the admixed configuration $K'$.
For details on how to find $\overline{E}\left(K\right)$ and $\overline{E}\left(K^{'}\right)$, see \cite[Section~3]{{Gaigetal:2024CV}}.
We would like to emphasize that the energy denominator (\ref{eq:BogO1}) is defined differently/opposite to the expressions of Feynman diagrams 
(see for example Fig.~\ref{VV_1}, Eqs. (\ref{eq:BogP}), (\ref{eq:BogQ})).

As we can see, the expression of the first type of valence-valence correlations  for the contribution $\Delta \mathcal{E}_0$ (see Table~\ref{tab:Implemen_VV1}) depends on the occupation number $w_m$ of subshell $m$ (number of electrons in the subshell $m$). 
This is due to the fact that the first type of valence-valence correlations is described by the VV$_1$ and VV$_2$ Feynman diagrams, whose angular part is expressed via the operator $\hat{N}$ of subshell occupation number (see Eqs. (\ref{eq:s-b_VV3}) and (\ref{eq:VV3-d})). Meanwhile, the second type of valence-valence correlations for the contribution $\Delta \mathcal{E}_0$  
described only by the VV$_2$ Feynman diagram 
 depends on the occupation number $w_m$ of subshell $m$ and
on the occupation number $w_n$ of subshell $n$ because the tensorial part of the diagram (see Eq. (\ref{eq:VV4-d})) act on these two subshells separately.

\begin{table*}
\begin{center}
\begin{tabular}{|lll|} \hline
Corrections & Slater integral & $k$ values\\ \hline  \hline
& & \\
$\underbrace{ 
- \; 4 \, \left[ k \right]  \mathcal{X}\left( k, \, m m, \, r s, \, n n \right)}_{\text{from $\text{VV}_{2}$ Feynman diagram}}$  \hspace{2cm}
&$\Delta \mathcal{R}^{k}(mm,nn)$ & $k\geq 0$\\
& & \\ \hline 
\end{tabular}
\end{center}
\caption{Expressions for Slater integrals $\Delta \mathcal{R}^{k}(mm,nn)$ (see  Eq. (\ref{eq:VVBogEnergy})) corresponding 
to the valence-valence $(n_{m} \ell_{m}) \; j_{m}^{w_m} \; (n_{n} \ell_{n}) \; j_{n}^{w_n} 
   \rightarrow (n_{m} \ell_{m})\, j_{m}^{w_m-2} \; (n_{n} \ell_{n}) \; j_{n}^{w_n}
	 \; (n_{r} \ell_{r}) \; j_{r} \; (n_{s} \ell_{s}) \; j_{s}$
corrections coming from the off diagional matrix element $\redmem{(n_{m} \ell_{m})\, j_{m}^{w_m} \, (n_{n} \ell_{n})\, j_{n}^{w_n}}{\, \widehat{{\cal H}}^{(2)}_{\text{Effective}} \,}{(n_{m} \ell_{m})\, j_{m}^{w_m-2} \, (n_{n} \ell_{n})\, j_{n}^{w_n+2} }$.} 
\label{tab:Implemen_VVc}
\end{table*}

The contribution of VV correlation in the second-order of the perturbation theory coming from off diagonal matrix element 
$\redmem{(n_{m} \ell_{m})\, j_{m}^{w_m} \, (n_{n} \ell_{n})\, j_{n}^{w_n}}{\, \widehat{{\cal H}}^{(2)}_{\text{Effective}} \,}{(n_{m} \ell_{m})\, j_{m}^{w_m-2} \, (n_{n} \ell_{n})\, j_{n}^{w_n+2} }$
is described by the diagram VV$_2$. Reformulation of the expressions of this diagram into the form suitable for the {\sc Grasp} gave these corrections only to the radial integral $\Delta \mathcal{R}^{k}(m m,nn)$ (see Table~\ref{tab:Implemen_VVc}). This formula is expressed via the quantity: 

\begin{equation}
\label{eq:BogX}
   \mathcal{X}\left(x, \; i j, \; i' j', \; i'' j''\right) 
  = \sum_{k,k'}
				  \left\{
    \begin{array}{ccc}
      k  & k' & x \\
      j_{i''} & j_{i} & j_{i'}
    \end{array} \right\}				  \left\{
    \begin{array}{ccc}
      k  & k' & x \\
      j_{j''} & j_{j} & j_{j'}
    \end{array} \right\}
\mathcal{S'}\left(kk', \; i j, \; i' j', \; i'' j''\right) ,
\end{equation}
where
\begin{equation}
\label{eq:BogSprim}
   \mathcal{S'}\left(kk', \; i j, \; i' j', \; i'' j''\right) 
  =
\mathcal{R}^{k}\left(i j, \; i' j'\right)  \; 
\mathcal{R}^{k'}\left(i' j', \; i'' j'' \right)  \; 
\mathcal{O}\left(K', K_1 K_2 \right)
\end{equation}
and
\begin{equation}
\label{eq:BogO2}
\mathcal{O}\left(K', K_1 K_2 \right)
=
\frac{1}{2} \left( {\frac{1}{\overline{E}\left(K' \right)-\overline{E}\left(K_1\right)}}
+{\frac{1}{\overline{E}\left(K' \right)-\overline{E}\left(K_2\right)}}
\right),
\end{equation}
where $\overline{E}\left(K_1\right)$ corresponds to the  averaged  energy of the configuration $K_1$ from bra function of off diagonal matrix element and  $\overline{E}\left(K_2\right)$ corresponds to the  averaged  energy of the configuration $K_1$ from ket function of off diagonal matrix element. For details on how to find them, see \cite[Section~3]{{Gaigetal:2024CV}}.

This theory in irreducible tensorial form is more suitable to be included in such version of the {\sc Grasp} which is based on configuration state function generators~\cite{grasp2023}. This is related to the fact that this version of the software package allows us to distinguish $F$, $F'$, and $G$ sets of orbitals very easily in the process of computing atomic data. In the following section we will present a test case of this implementation.

\section{Calculation of core-valence, core, core-core and valence-valence with a new approach}
\label{Sec:Calculations}
In the present work, the method
based on the Rayleigh-Schr\"odinger perturbation theory in an irreducible tensorial form \cite{Gaigetal:2024CV,Gaigetal:2024C,Gaigetal:2024CC}
is extended to include VV correlations in the computations. 
Firstly, we present the results when only VV correlations are included in a regular way 
and using the stationary second-order Rayleigh-Schr\"odinger many-body perturbation theory in an irreducible tensorial form.
For this purpose, we computed 105 lowest energy levels of the $\mathrm{4s^24p^2}$, $\mathrm{4p^4}$, 
$\mathrm{4s^24p\{4d,4f,5s,5p,5d,6s,6p\}}$, $\mathrm{4s4p^3}$, and 
$\mathrm{4s4p^2\{4d,5s\}}$ configurations of the Se~III.
Furthemore, we performed calculations using the regular way and the RSMBPT method when CV, C, CC and VV correlations
are included. To reduce the computational 
resources, these calculations are performed for Se~III energy levels of the even configurations 
with $J$ = 0 and the levels of the odd configurations with $J$ = 1. 
The radial wave functions for all above mentioned computations are taken from the earlier study \cite{Se_Ge_like}. 
The multireference (MR) set in the present calculations consists of the $\mathrm{4s^24p^2}$, $\mathrm{4p^4}$, 
$\mathrm{4s^24p\{4f,5p,6p\}}$, $\mathrm{4s4p^2\{4d,5s\}}$ even and 
$\mathrm{4s^24p\{4d,5s,5d,6s\}}$, $\mathrm{4s4p^3}$ odd configurations. 
It should be noted that the MR set in the present computations consists 
of orbitals with few different principal quantum numbers, namely with $n$, $n+1$, and $n+2$ (where $n$=4).
Such calculations are carried out using both the regular RCI and the RCI (RSMBPT) methods.
The RCI calculations are performed, including the Coulomb and
Breit interactions and leading quantum electrodynamic effects – the vacuum polarization and
self-energy corrections. Meanwhile, the estimation of correlations was done using the stationary
second-order Rayleigh-Schr\"odinger many-body perturbation theory in an irreducible tensorial
form for the Coulomb interaction. 
The calculations using both the regular RCI and RCI (RSMBPT) methods were performed including only CSFs that have non-zero
matrix elements in the sets of spin-angular integration with the CSFs belonging to the configurations
in the MR.

\subsection{Computational schemes}

Regular RCI computations including VV correlations are marked as \textbf{VV RCI}.
In this computational scheme single-double (SD) substitutions are allowed from 
the 4s, $\mathrm{4p_-}$, 4p, $\mathrm{4d_-}$, 4d, $\mathrm{4f_-}$, 4f,
5s, $\mathrm{5p_-}$, 5p, $\mathrm{5d_-}$, 5d, 6s, $\mathrm{6p_-}$, 6p valence orbitals of the MR set
to orbital set (OS) $OS_1$=\{7s,$\mathrm{7p_-}$,7p,$\mathrm{6d_-}$,6d,$\mathrm{5f_-}$,5f,$\mathrm{5g_-}$,5g\}, ..., 
$OS_4$=\{10s,$\mathrm{10p_-}$,10p,$\mathrm{9d_-}$,9d,$\mathrm{8f_-}$,8f,$\mathrm{8g_-}$,8g\}.
In \textbf{CV+C+CC+VV RCI} computational scheme 
SD substitutions are allowed from the valence orbitals (mentioned above)
and from the 3s, $\mathrm{3p_-}$, 3p, $\mathrm{3d_-}$ and 3d core orbitals to orbital set $OS_1$, ..., $OS_4$.

The CSF space in the computations using the RSMBPT method is divided into three sets: $F$, $F'$ and $G$ 
(see Ref. \cite{Gaigetal:2024CV} for details). 
The 1s, 2s, $\mathrm{2p_-}$, 2p, 
3s, $\mathrm{3p_-}$, 3p, $\mathrm{3d_-}$ and 3d subshells are defined as core subshells 
(that correspond to $F$ set), 
4s, $\mathrm{4p_-}$, 4p, $\mathrm{4d_-}$, 4d, $\mathrm{4f_-}$, 4f,
5s, $\mathrm{5p_-}$, 5p, $\mathrm{5d_-}$, 5d, 6s, $\mathrm{6p_-}$, 6p as valence subshells 
(that correspond to $F'$ set), and subshells belonging to $OS_1$, ..., $OS_4$ as virtual ones 
(that correspond to $G$ set). 
In the calculations when only VV correlations are included all core subshells are inactive, 
in case including CV, C, CC and VV correlations the 1s, 2s, $\mathrm{2p_-}$ and 2p subshells are defined as inactive core subshells, 
3s, $\mathrm{3p_-}$, 3p, $\mathrm{3d_-}$ and 3d subshells are defined as active core subshells.
This space distribution is consistent with regular {\sc Grasp}2018 calculations
and allows the use of a combination of RCI and RSMBPT methods.

The RSMBPT calculation procedure is analogous to that used in previous research \cite{Gaigetal:2024CV,Gaigetal:2024C,Gaigetal:2024CC}.
The contribution of each $K'$ configuration for CSF 
for which energy needs to be calculated according to Rayleigh-Schr\"odinger perturbation theory 
in an irreducible tensorial form is computed according 
to the Eq. (22) of Ref. \cite{Gaigetal:2024CV} (for CV correlations), Eq. (6) of Ref. \cite{Gaigetal:2024C} (for C correlations),
Eq. (26) of Ref. \cite{Gaigetal:2024CC} (for CC correlations), and Eq. (\ref{eq:BogEnergy_PT}) (for VV correlations).
$K'$ configurations are sorted in descending order according to the impact of the correlations for each level.
Further, $K'$ configurations are selected by the VV (for calculations when only VV correlations included) 
or CV, C, CC and VV (for calculations when all correlations are included) correlations impact
with the specified fraction (expressed in the percentage: 95, 99, 99.5, 99.95 and 100\%) 
of the total correlations contribution, and RCI computations are performed including them.
It should be noted that the program gives the contribution of the correlations of $K'$ configuration 
with a value greater than {\tt 1.0E-11}. Contributions of smaller magnitudes are neglected.
The C correlations (Eq. (3) of Ref. \cite{Gaigetal:2024C}), CV correlations (Eq. (4) of Ref. \cite{Gaigetal:2024C}) and 
VV correlations (Eqs. (\ref{eq:remaining_VV-a}) and (\ref{eq:remaining_VV-b})) which are not included with RSMBPT method, 
were added to RCI calculations in a regular way.
Results that include VV correlations according to the RSMBPT method are marked as \textbf{VV RCI (RSMBPT)}.
Results including CV, C, CC and VV correlations according to the RSMBPT method are marked as \textbf{CV+C+CC+VV RCI (RSMBPT)}.

\subsection{Results}
\subsubsection{Valence-valence correlations}

Table \ref{comp1} presents a comparison of the total energies 
from regular RCI (\textbf{VV RCI}) and from RSMBPT (\textbf{VV RCI (RSMBPT)}) calculations 
for 105 energy levels of the Se~III. 
In the table total energies from \textbf{VV RCI} calculations and the energy differences between 
\textbf{VV RCI (RSMBPT)} and \textbf{VV RCI} calculations are given. 
The last line of the table gives the number of CSFs ($N_{CSFs}$) for each computational scheme.
As seen from the table, including the most significant VV correlations by increasing the
amount of these correlations (95, 99, 99.5, 99.95 and 100\%), the results converge 
to the regular {\sc Grasp}2018 results and in the case of '100\%' reproduce them.
In the case of '100\%' for the majority of levels the agreement is excellent, the discrepancy
between \textbf{VV RCI (RSMBPT)} and \textbf{VV RCI} is very small and reach only up to 4.9E-05 a.u. ( or 0.000002\%).

Table \ref{comp2} compares the energy levels from regular RCI and RSMBPT computations when only VV correlations are included. 
It is seen from the table, that by adding the most important $K'$ configurations of VV correlations step by step, 
the \textbf{VV RCI (RSMBPT)} results converge to the results of regular {\sc Grasp}2018 calculations. 
The largest difference in the case of '100\%' is 10.17 cm$^{-1}$, and for the most levels the
difference between \textbf{VV RCI (RSMBPT)} and \textbf{VV RCI} calculations do not reach 1 cm$^{-1}$.

\subsubsection{Core-valence, core, core-core, and valence-valence correlations}
Table \ref{comp3} displays the total energies from \textbf{CV+C+CC+VV RCI} calculations and the energy differences between 
\textbf{CV+C+CC+VV RCI (RSMBPT)} and \textbf{CV+C+CC+VV RCI} calculations.
For these computations including all electron correlations, as it mentioned above,
only energy levels of the even configurations 
with $J$ = 0 and the levels of the odd configurations with $J$ = 1 are investigated.
The calculations using the RSMBPT method are carried out in five steps, including 
95, 99, 99.5, 99.95 and 100\% of CV, C, CC and VV correlations. 
The results of the \textbf{CV+C+CC+VV RCI (RSMBPT)} in the case of '100\%' 
reproduce the regular {\sc Grasp}2018 results. The largest difference between these results is 2.63E-05 a.u. ( or 0.0000011\%)

In Table \ref{comp4} the energy levels 
from calculations using the RSMBPT method (\textbf{CV+C+CC+VV RCI (RSMBPT)}) are compared. 
In the last line of the table the root-mean-square (rms) with results of regular 
{\sc Grasp}2018 calculations (\textbf{CV+C+CC+VV RCI}) are given. 
When the most important $K'$ configurations of CV, C, CC and VV correlations are progressively included in the calculations
the results converge to the results of the regular {\sc Grasp}2018 calculations 
and agree very well with them in the case of '100\%'. 
The rms with \textbf{CV+C+CC+VV RCI} results is 1.90 cm$^{-1}$, and the largest difference between the results 
of these calculations is only up to 4.84 cm$^{-1}$.
Omitting CSFs with smaller impact of the CV, C, CC and VV correlations, 
the energy levels are still in good agreement with the regular computations but the space of CSFs decreases significantly. 
For example, in the '99.5\%' case, the rms with \textbf{CV+C+CC+VV RCI} results is 94.63 cm$^{-1}$, 
while the space of CSFs decreases by a factor of 1.8 comparing to the space in \textbf{CV+C+CC+VV RCI} computations.

The presented results demonstrate that the RCI (RSMBPT) method works very well even when the MR set consists 
of orbitals with few different principal quantum numbers, namely with $n$, $n+1$, and $n+2$ (where $n$=4).
The results obtained using the RCI (RSMBPT) method 
agree very well with the regular {\sc Grasp}2018 results in both computations 
when only VV and when all electron correlations are included. 
In addition, the RCI (RSMBPT) method works very well when MR set consists from configurations 
with any orbital quantum numbers $l$, and with any shells occupation. 
Such studies have already been carried out in Refs. \cite{Gaigetal:2024C,Gaigetal:2024CC}, 
and confirmed in the present work.

{\scriptsize
\begin{longtable}{r l r r r r r r}
\caption{\label{comp1} The total energies (in a.u.) from \textbf{VV RCI} calculations and differences (in a.u.) between \textbf{VV RCI (RSMBPT)} and \textbf{VV RCI} 
 energies ($\Delta E_{\textbf{(VV~RCI (RSMBPT))-(VV~RCI)}}$) for Se~III are given when VV correlations are included in the computations.}\\
\hline\hline
\multicolumn{1}{c}{\multirow{2}{*}{No.}} & \multicolumn{1}{c}{\multirow{2}{*}{State}} & \multicolumn{1}{c}{\multirow{2}{*}{\textbf{VV RCI}}}& \multicolumn{5}{c}{$\Delta E_{\textbf{(VV~RCI (RSMBPT))-(VV~RCI)}}$} \\
\cline{4-8}
&&& \multicolumn{1}{c}{95\%} & \multicolumn{1}{c}{99\%} & \multicolumn{1}{c}{99.5\%} & \multicolumn{1}{c}{99.95\%} & \multicolumn{1}{c}{100\%} \\
\hline
\noalign{\smallskip}
\endfirsthead
\caption{Continued.}\\
\hline\hline
\multicolumn{1}{c}{\multirow{2}{*}{No.}} & \multicolumn{1}{c}{\multirow{2}{*}{State}} & \multicolumn{1}{c}{\multirow{2}{*}{\textbf{VV RCI}}}& \multicolumn{5}{c}{$\Delta E_{\textbf{(VV~RCI (RSMBPT))-(VV~RCI)}}$} \\
\cline{4-8}
&&& \multicolumn{1}{c}{95\%} & \multicolumn{1}{c}{99\%} & \multicolumn{1}{c}{99.5\%} & \multicolumn{1}{c}{99.95\%} & \multicolumn{1}{c}{100\%} \\
\hline
\noalign{\smallskip}
\endhead
\hline
\hline
\endfoot
\noalign{\smallskip}
   1& $\mathrm{4s^24p^2~^3P_0}$       & -2425.6827502 & 0.0001895 & 0.0000309 & 0.0000202 & 0.0000120 & 0.0000027  \\
   2& $\mathrm{4s^24p^2~^3P_1}$       & -2425.6753106 & 0.0001577 & 0.0000308 & 0.0000165 & 0.0000038 & 0.0000020  \\
   3& $\mathrm{4s^24p^2~^3P_2}$       & -2425.6656787 & 0.0001364 & 0.0000256 & 0.0000117 & 0.0000025 & 0.0000021  \\
   4& $\mathrm{4s^24p^2~^1D_2}$       & -2425.6233754 & 0.0002463 & 0.0000542 & 0.0000263 & 0.0000042 & 0.0000031  \\
   5& $\mathrm{4s^24p^2~^1S_0}$       & -2425.5540801 & 0.0002955 & 0.0000689 & 0.0000330 & 0.0000252 & 0.0000014  \\
   6& $\mathrm{4s4p^3~^5S^o_2}$       & -2425.3865441 & 0.0000772 & 0.0000170 & 0.0000081 & 0.0000010 & 0.0000000  \\
   7& $\mathrm{4s4p^3~^3D^o_1}$       & -2425.2728556 & 0.0001927 & 0.0000496 & 0.0000340 & 0.0000121 & 0.0000071  \\
   8& $\mathrm{4s4p^3~^3D^o_2}$       & -2425.2722502 & 0.0001830 & 0.0000383 & 0.0000222 & 0.0000084 & 0.0000072  \\
   9& $\mathrm{4s4p^3~^3D^o_3}$       & -2425.2696375 & 0.0001987 & 0.0000466 & 0.0000291 & 0.0000100 & 0.0000058  \\
  10& $\mathrm{4s4p^3~^3P^o_2}$       & -2425.2080103 & 0.0002059 & 0.0000384 & 0.0000211 & 0.0000044 & 0.0000027  \\
  11& $\mathrm{4s4p^3~^3P^o_0}$       & -2425.2077293 & 0.0002487 & 0.0000750 & 0.0000402 & 0.0000122 & 0.0000010  \\
  12& $\mathrm{4s4p^3~^3P^o_1}$       & -2425.2077069 & 0.0003106 & 0.0000564 & 0.0000395 & 0.0000102 & 0.0000041  \\
  13& $\mathrm{4s^24p4d~^1D^o_2}$     & -2425.1795605 & 0.0001731 & 0.0000333 & 0.0000215 & 0.0000076 & 0.0000062  \\
  14& $\mathrm{4s^24p4d~^3F^o_2}$     & -2425.1264631 & 0.0002440 & 0.0000663 & 0.0000457 & 0.0000271 & 0.0000239  \\
  15& $\mathrm{4s^24p4d~^3F^o_3}$     & -2425.1204839 & 0.0002133 & 0.0000474 & 0.0000303 & 0.0000125 & 0.0000092  \\
  16& $\mathrm{4s^24p5s~^3P^o_0}$     & -2425.1130178 & 0.0002397 & 0.0000860 & 0.0000682 & 0.0000445 & 0.0000342  \\
  17& $\mathrm{4s^24p5s~^3P^o_1}$     & -2425.1107765 & 0.0001492 & 0.0000293 & 0.0000167 & 0.0000036 & 0.0000020  \\
  18& $\mathrm{4s^24p4d~^3F^o_4}$     & -2425.1102801 & 0.0003001 & 0.0000771 & 0.0000460 & 0.0000069 & 0.0000002  \\
  19& $\mathrm{4s^24p5s~^3P^o_2}$     & -2425.0952711 & 0.0001736 & 0.0000404 & 0.0000244 & 0.0000038 & 0.0000006  \\
  20& $\mathrm{4s^24p5s~^1P^o_1}$     & -2425.0877335 & 0.0001506 & 0.0000248 & 0.0000145 & 0.0000022 & 0.0000010  \\
  21& $\mathrm{4s^24p4d~^1P^o_1}$     & -2425.0559576 & 0.0002530 & 0.0000591 & 0.0000391 & 0.0000191 & 0.0000164  \\
  22& $\mathrm{4s^24p4d~^3P^o_2}$     & -2425.0487287 & 0.0003104 & 0.0000608 & 0.0000333 & 0.0000077 & 0.0000044  \\
  23& $\mathrm{4s^24p4d~^3D^o_1}$     & -2425.0469453 & 0.0002286 & 0.0000543 & 0.0000329 & 0.0000141 & 0.0000097  \\
  24& $\mathrm{4s^24p4d~^3P^o_1}$     & -2425.0367313 & 0.0002198 & 0.0000538 & 0.0000344 & 0.0000169 & 0.0000150  \\
  25& $\mathrm{4s^24p4d~^3D^o_3}$     & -2425.0365571 & 0.0002450 & 0.0000468 & 0.0000313 & 0.0000119 & 0.0000086  \\
  26& $\mathrm{4s^24p4d~^3P^o_0}$     & -2425.0361679 & 0.0003669 & 0.0001383 & 0.0001016 & 0.0000676 & 0.0000490  \\
  27& $\mathrm{4s^24p4d~^3D^o_2}$     & -2425.0342497 & 0.0002600 & 0.0000569 & 0.0000399 & 0.0000179 & 0.0000155  \\
  28& $\mathrm{4s4p^3~^3S^o_1}$       & -2425.0324298 & 0.0002367 & 0.0000602 & 0.0000432 & 0.0000116 & 0.0000077  \\
  29& $\mathrm{4s^24p4d~^1F^o_3}$     & -2425.0072526 & 0.0003419 & 0.0000807 & 0.0000468 & 0.0000253 & 0.0000191  \\
  30& $\mathrm{4s^24p5p~^1P_1}$       & -2425.0038073 & 0.0001521 & 0.0000368 & 0.0000242 & 0.0000108 & 0.0000006  \\
  31& $\mathrm{4s4p^3~^1D^o_2}$       & -2425.0023171 & 0.0002710 & 0.0000591 & 0.0000397 & 0.0000162 & 0.0000143  \\
  32& $\mathrm{4s^24p5p~^3D_1}$       & -2424.9927206 & 0.0001462 & 0.0000408 & 0.0000220 & 0.0000082 & 0.0000005  \\
  33& $\mathrm{4s^24p5p~^3D_2}$       & -2424.9911607 & 0.0001331 & 0.0000320 & 0.0000181 & 0.0000038 & 0.0000006  \\
  34& $\mathrm{4s^24p5p~^3P_0}$       & -2424.9849192 & 0.0001334 & 0.0000262 & 0.0000187 & 0.0000081 & 0.0000006  \\
  35& $\mathrm{4s^24p5p~^3P_1}$       & -2424.9785409 & 0.0001262 & 0.0000312 & 0.0000184 & 0.0000065 & 0.0000003  \\
  36& $\mathrm{4s^24p5p~^3D_3}$       & -2424.9774657 & 0.0002120 & 0.0000616 & 0.0000335 & 0.0000115 & 0.0000001  \\
  37& $\mathrm{4s^24p5p~^3P_2}$       & -2424.9716950 & 0.0001131 & 0.0000241 & 0.0000139 & 0.0000033 & 0.0000005  \\
  38& $\mathrm{4s^24p5p~^3S_1}$       & -2424.9652235 & 0.0001870 & 0.0000613 & 0.0000375 & 0.0000163 & 0.0000001  \\
  39& $\mathrm{4s4p^3~^1P^o_1}$       & -2424.9597596 & 0.0003101 & 0.0000696 & 0.0000445 & 0.0000153 & 0.0000113  \\
  40& $\mathrm{4s^24p5p~^1D_2}$       & -2424.9561445 & 0.0001302 & 0.0000278 & 0.0000162 & 0.0000037 & 0.0000011  \\
  41& $\mathrm{4s^24p5p~^1S_0}$       & -2424.9308286 & 0.0001819 & 0.0000432 & 0.0000246 & 0.0000088 & 0.0000005  \\
  42& $\mathrm{4s^24p6s~^3P^o_0}$     & -2424.8379021 & 0.0001859 & 0.0000590 & 0.0000462 & 0.0000302 & 0.0000232  \\
  43& $\mathrm{4s^24p6s~^3P^o_1}$     & -2424.8368552 & 0.0001271 & 0.0000222 & 0.0000120 & 0.0000035 & 0.0000028  \\
  44& $\mathrm{4s^24p5d~^3F^o_2}$     & -2424.8330064 & 0.0001443 & 0.0000349 & 0.0000248 & 0.0000091 & 0.0000067  \\
  45& $\mathrm{4s^24p5d~^3F^o_3}$     & -2424.8272709 & 0.0001368 & 0.0000272 & 0.0000143 & 0.0000040 & 0.0000018  \\
  46& $\mathrm{4s^24p4f~^1F_3}$       & -2424.8271943 & 0.0001997 & 0.0000591 & 0.0000337 & 0.0000124 & 0.0000065  \\
  47& $\mathrm{4s^24p4f~^3F_3}$       & -2424.8257288 & 0.0001971 & 0.0000364 & 0.0000256 & 0.0000068 & 0.0000016  \\
  48& $\mathrm{4s^24p4f~^3F_2}$       & -2424.8249661 & 0.0001671 & 0.0000318 & 0.0000242 & 0.0000101 & 0.0000091  \\
  49& $\mathrm{4s^24p5d~^3D_2}$       & -2424.8245284 & 0.0001321 & 0.0000208 & 0.0000124 & 0.0000027 & 0.0000015  \\
  50& $\mathrm{4s^24p4f~^3F_4}$       & -2424.8245239 & 0.0001761 & 0.0000459 & 0.0000235 & 0.0000042 & 0.0000003  \\
  51& $\mathrm{4s^24p5d~^3D^o_1}$     & -2424.8207738 & 0.0001337 & 0.0000216 & 0.0000144 & 0.0000070 & 0.0000056  \\
  52& $\mathrm{4s^24p6s~^3P^o_2}$     & -2424.8187223 & 0.0001537 & 0.0000313 & 0.0000205 & 0.0000038 & 0.0000012  \\
  53& $\mathrm{4s4p^2~^4P~4d~^3P_2}$  & -2424.8178397 & 0.0001080 & 0.0000235 & 0.0000111 & 0.0000047 & 0.0000036  \\
  54& $\mathrm{4s^24p6s~^1P^o_1}$     & -2424.8156645 & 0.0001222 & 0.0000183 & 0.0000103 & 0.0000027 & 0.0000018  \\
  55& $\mathrm{4s^24p5d~^3F^o_4}$     & -2424.8138892 & 0.0001600 & 0.0000341 & 0.0000198 & 0.0000033 & 0.0000001  \\
  56& $\mathrm{4s4p^2~^4P~4d~^5F_1}$  & -2424.8119309 & 0.0001347 & 0.0000325 & 0.0000212 & 0.0000086 & 0.0000051  \\
  57& $\mathrm{4s^24p4f~^3G_3}$       & -2424.8105949 & 0.0001999 & 0.0000552 & 0.0000368 & 0.0000159 & 0.0000097  \\
  58& $\mathrm{4s^24p5d~^1D^o_2}$     & -2424.8103478 & 0.0001300 & 0.0000246 & 0.0000162 & 0.0000064 & 0.0000050  \\
  59& $\mathrm{4s4p^2~^4P~4d~^5F_2}$  & -2424.8097339 & 0.0001026 & 0.0000232 & 0.0000138 & 0.0000050 & 0.0000043  \\
  60& $\mathrm{4s^24p4f~^3G_4}$       & -2424.8091362 & 0.0001252 & 0.0000385 & 0.0000273 & 0.0000053 & 0.0000010  \\
  61& $\mathrm{4s4p^2~^4P~4d~^3P_1}$  & -2424.8091238 & 0.0001718 & 0.0000372 & 0.0000161 & 0.0000050 & 0.0000025  \\
  62& $\mathrm{4s^24p5d~^3D^o_3}$     & -2424.8072808 & 0.0001359 & 0.0000257 & 0.0000171 & 0.0000053 & 0.0000035  \\
  63& $\mathrm{4s4p^2~^4P~4d~^5F_3}$  & -2424.8063898 & 0.0001154 & 0.0000269 & 0.0000180 & 0.0000079 & 0.0000031  \\
  64& $\mathrm{4s4p^2~^4P~4d~^3P_0}$  & -2424.8060102 & 0.0001349 & 0.0000239 & 0.0000152 & 0.0000052 & 0.0000009  \\
  65& $\mathrm{4s^24p4f~^3G_5}$       & -2424.8058236 & 0.0002456 & 0.0000808 & 0.0000445 & 0.0000118 & 0.0000003  \\
  66& $\mathrm{4s^24p5d~^3P^o_2}$     & -2424.8038729 & 0.0001341 & 0.0000273 & 0.0000168 & 0.0000060 & 0.0000045  \\
  67& $\mathrm{4s^24p4f~^3D_3}$       & -2424.8033484 & 0.0001971 & 0.0000414 & 0.0000256 & 0.0000070 & 0.0000009  \\
  68& $\mathrm{4s^24p5d~^3P^o_1}$     & -2424.8029622 & 0.0001490 & 0.0000402 & 0.0000253 & 0.0000096 & 0.0000074  \\
  69& $\mathrm{4s^24p4f~^3D_2}$       & -2424.8021879 & 0.0001782 & 0.0000370 & 0.0000246 & 0.0000089 & 0.0000070  \\
  70& $\mathrm{4s^24p5d~^3P^o_0}$     & -2424.8020585 & 0.0001877 & 0.0000501 & 0.0000346 & 0.0000204 & 0.0000130  \\
  71& $\mathrm{4s4p^2~^4P~4d~^5F _4}$ & -2424.8016324 & 0.0001099 & 0.0000219 & 0.0000143 & 0.0000041 & 0.0000016  \\
  72& $\mathrm{4s^24p4f~^1G_4}$       & -2424.7995880 & 0.0002058 & 0.0000625 & 0.0000403 & 0.0000074 & 0.0000022  \\
  73& $\mathrm{4s^24p4f~^3D_1}$       & -2424.7985206 & 0.0002064 & 0.0000403 & 0.0000283 & 0.0000085 & 0.0000044  \\
  74& $\mathrm{4s^24p4f~^1D_2}$       & -2424.7969189 & 0.0001596 & 0.0000366 & 0.0000215 & 0.0000024 & 0.0000007  \\
  75& $\mathrm{4s4p^2~^4P~4d~^5F_5}$  & -2424.7951151 & 0.0001537 & 0.0000455 & 0.0000311 & 0.0000127 & 0.0000005  \\
  76& $\mathrm{4s^24p5d~^1F^o_3}$     & -2424.7928576 & 0.0002132 & 0.0000466 & 0.0000275 & 0.0000127 & 0.0000085  \\
  77& $\mathrm{4s^24p6p~^3D_1}$       & -2424.7927791 & 0.0001832 & 0.0000451 & 0.0000280 & 0.0000100 & 0.0000013  \\
  78& $\mathrm{4s^24p6p~^3P_1}$       & -2424.7892989 & 0.0001538 & 0.0000366 & 0.0000209 & 0.0000054 & 0.0000006  \\
  79& $\mathrm{4s^24p6p~^3D_2}$       & -2424.7880545 & 0.0001348 & 0.0000276 & 0.0000164 & 0.0000021 & 0.0000006  \\
  80& $\mathrm{4s^24p5d~^1P^o_1}$     & -2424.7876585 & 0.0001892 & 0.0000410 & 0.0000291 & 0.0000106 & 0.0000076  \\
  81& $\mathrm{4s^24p6p~^3P_0}$       & -2424.7874870 & 0.0001368 & 0.0000279 & 0.0000208 & 0.0000065 & 0.0000005  \\
  82& $\mathrm{4s4p^2~^4P~4d~^5D_0}$  & -2424.7825283 & 0.0001321 & 0.0000315 & 0.0000210 & 0.0000072 & 0.0000034  \\
  83& $\mathrm{4s4p^2~^4P~4d~^5D_1}$  & -2424.7823026 & 0.0001235 & 0.0000296 & 0.0000195 & 0.0000072 & 0.0000036  \\
  84& $\mathrm{4s4p^2~^4P~4d~^5D_2}$  & -2424.7815098 & 0.0001067 & 0.0000222 & 0.0000145 & 0.0000044 & 0.0000033  \\
  85& $\mathrm{4s4p^2~^4P~4d~^5D_3}$  & -2424.7799579 & 0.0001435 & 0.0000322 & 0.0000226 & 0.0000093 & 0.0000024  \\
  86& $\mathrm{4s4p^2~^4P~4d~^5D_4}$  & -2424.7770902 & 0.0001382 & 0.0000297 & 0.0000208 & 0.0000064 & 0.0000013  \\
  87& $\mathrm{4s^24p6p~^1P_1}$       & -2424.7754567 & 0.0001410 & 0.0000353 & 0.0000209 & 0.0000055 & 0.0000008  \\
  88& $\mathrm{4s4p^2~^4P~5s~^5P_1}$  & -2424.7740592 & 0.0001216 & 0.0000344 & 0.0000226 & 0.0000044 & 0.0000010  \\
  89& $\mathrm{4s^24p6p~^3P_2}$       & -2424.7729862 & 0.0001384 & 0.0000327 & 0.0000169 & 0.0000033 & 0.0000012  \\
  90& $\mathrm{4s^24p6p~^3D_3}$       & -2424.7719517 & 0.0001989 & 0.0000506 & 0.0000271 & 0.0000077 & 0.0000003  \\
  91& $\mathrm{4p^4~^1D_2}$           & -2424.7709312 & 0.0001722 & 0.0000698 & 0.0000383 & 0.0000056 & 0.0000031  \\
  92& $\mathrm{4s^24p6p~^3S_1}$       & -2424.7681892 & 0.0001537 & 0.0000385 & 0.0000240 & 0.0000091 & 0.0000002  \\
  93& $\mathrm{4s4p^2~^4P~5s~^5P_2}$  & -2424.7675863 & 0.0000780 & 0.0000215 & 0.0000159 & 0.0000018 & 0.0000011  \\
  94& $\mathrm{4s^24p6p~^1D_2}$       & -2424.7652980 & 0.0001359 & 0.0000298 & 0.0000163 & 0.0000027 & 0.0000009  \\
  95& $\mathrm{4s4p^2~^4P~5s~^5P_3}$  & -2424.7593175 & 0.0000961 & 0.0000229 & 0.0000161 & 0.0000048 & 0.0000008  \\
  96& $\mathrm{4s^24p6p~^1S_0}$       & -2424.7562436 & 0.0001812 & 0.0000421 & 0.0000291 & 0.0000086 & 0.0000007  \\
  97& $\mathrm{4s4p^2~^4P~4d~^3F_2}$  & -2424.7533416 & 0.0002147 & 0.0001140 & 0.0000915 & 0.0000102 & 0.0000055  \\
  98& $\mathrm{4s4p^2~^4P~4d~^3F_3}$  & -2424.7483267 & 0.0002850 & 0.0001041 & 0.0000916 & 0.0000312 & 0.0000047  \\
  99& $\mathrm{4s4p^2~^4P~4d~^3F_4}$  & -2424.7411003 & 0.0002728 & 0.0000810 & 0.0000493 & 0.0000112 & 0.0000036  \\
 100& $\mathrm{4s4p^2~^4P~5s~^3P_0}$  & -2424.7382269 & 0.0002030 & 0.0000621 & 0.0000522 & 0.0000166 & 0.0000026  \\
 101& $\mathrm{4s4p^2~^4P~5s~^3P_1}$  & -2424.7330497 & 0.0002147 & 0.0000900 & 0.0000591 & 0.0000072 & 0.0000014  \\
 102& $\mathrm{4s4p^2~^4P~4d~^5P_3}$  & -2424.7304806 & 0.0001247 & 0.0000255 & 0.0000183 & 0.0000076 & 0.0000014  \\
 103& $\mathrm{4s4p^2~^4P~4d~^5P_2}$  & -2424.7277161 & 0.0000887 & 0.0000204 & 0.0000140 & 0.0000049 & 0.0000037  \\
 104& $\mathrm{4s4p^2~^4P~4d~^5P_1}$  & -2424.7255078 & 0.0001077 & 0.0000287 & 0.0000208 & 0.0000104 & 0.0000052  \\
 105& $\mathrm{4s4p^2~^4P~5s~^3P_2}$  & -2424.7232999 & 0.0001736 & 0.0000560 & 0.0000297 & 0.0000030 & 0.0000020  \\
\hline
\noalign{\smallskip}
\multicolumn{2}{l}{$N_{CSFs}$} &    285503&    176181&	219451&	232906&	259552& 274526  \\
\end{longtable}
}

{\scriptsize
\begin{longtable}{r l r r r r r r}
\caption{\label{comp2} The energy levels (in cm$^{-1}$) and differences (in cm$^{-1}$) between \textbf{VV RCI (RSMBPT)} and \textbf{VV RCI} 
 energies ($\Delta E_{\textbf{(VV~RCI (RSMBPT))-(VV~RCI)}}$) for Se~III are given when VV correlations are included in the computations.}\\
\hline\hline
\multicolumn{1}{c}{\multirow{2}{*}{No.}} & \multicolumn{1}{c}{\multirow{2}{*}{State}} & \multicolumn{1}{c}{\multirow{2}{*}{\textbf{VV RCI}}}& \multicolumn{5}{c}{$\Delta E_{\textbf{(VV~RCI (RSMBPT))-(VV~RCI)}}$} \\
\cline{4-8}
&&& \multicolumn{1}{c}{95\%} & \multicolumn{1}{c}{99\%} & \multicolumn{1}{c}{99.5\%} & \multicolumn{1}{c}{99.95\%} & \multicolumn{1}{c}{100\%} \\
\hline
\noalign{\smallskip}
\endfirsthead
\caption{Continued.}\\
\hline\hline
\multicolumn{1}{c}{\multirow{2}{*}{No.}} & \multicolumn{1}{c}{\multirow{2}{*}{State}} & \multicolumn{1}{c}{\multirow{2}{*}{\textbf{VV RCI}}}& \multicolumn{5}{c}{$\Delta E_{\textbf{(VV~RCI (RSMBPT))-(VV~RCI)}}$} \\
\cline{4-8}
&&& \multicolumn{1}{c}{95\%} & \multicolumn{1}{c}{99\%} & \multicolumn{1}{c}{99.5\%} & \multicolumn{1}{c}{99.95\%} & \multicolumn{1}{c}{100\%} \\
\hline
\noalign{\smallskip}
\endhead
\hline
\hline
\endfoot
\noalign{\smallskip}
   1& $\mathrm{4s^24p^2~^3P_0}$       & 0.00      & 0.00   & 0.00  & 0.00  & 0.00  & 0.00   \\
   2& $\mathrm{4s^24p^2~^3P_1}$       & 1632.80   & $-$6.98  & $-$0.01 & $-$0.81 & $-$1.82 & $-$0.15  \\
   3& $\mathrm{4s^24p^2~^3P_2}$       & 3746.75   & $-$11.65 & $-$1.15 & $-$1.86 & $-$2.08 & $-$0.12  \\
   4& $\mathrm{4s^24p^2~^1D_2}$       & 13031.27  & 12.45  & 5.10  & 1.34  & $-$1.73 & 0.07   \\
   5& $\mathrm{4s^24p^2~^1S_0}$       & 28239.83  & 23.24  & 8.34  & 2.81  & 2.87  & $-$0.29  \\
   6& $\mathrm{4s4p^3~^5S^o_2}$       & 65009.71  & $-$24.64 & $-$3.03 & $-$2.62 & $-$2.42 & $-$0.59  \\
   7& $\mathrm{4s4p^3~^3D^o_1}$       & 89961.46  & 0.70   & 4.12  & 3.05  & 0.03  & 0.97   \\
   8& $\mathrm{4s4p^3~^3D^o_2}$       & 90094.34  & $-$1.44  & 1.63  & 0.43  & $-$0.80 & 0.98   \\
   9& $\mathrm{4s4p^3~^3D^o_3}$       & 90667.75  & 2.03   & 3.46  & 1.97  & $-$0.45 & 0.68   \\
  10& $\mathrm{4s4p^3~^3P^o_2}$       & 104193.36 & 3.60   & 1.65  & 0.22  & $-$1.68 & 0.00   \\
  11& $\mathrm{4s4p^3~^3P^o_0}$       & 104255.04 & 12.99  & 9.68  & 4.40  & 0.02  & $-$0.38  \\
  12& $\mathrm{4s4p^3~^3P^o_1}$       & 104259.96 & 26.57  & 5.60  & 4.24  & $-$0.42 & 0.31   \\
  13& $\mathrm{4s^24p4d~^1D^o_2}$     & 110437.37 & $-$3.60  & 0.53  & 0.29  & $-$0.96 & 0.77   \\
  14& $\mathrm{4s^24p4d~^3F^o_2}$     & 122090.91 & 11.96  & 7.78  & 5.61  & 3.30  & 4.65   \\
  15& $\mathrm{4s^24p4d~^3F^o_3}$     & 123403.18 & 5.23   & 3.64  & 2.24  & 0.11  & 1.44   \\
  16& $\mathrm{4s^24p5s~^3P^o_0}$     & 125041.82 & 11.00  & 12.09 & 10.54 & 7.12  & 6.89   \\
  17& $\mathrm{4s^24p5s~^3P^o_1}$     & 125533.72 & $-$8.85  & $-$0.35 & $-$0.76 & $-$1.85 & $-$0.15  \\
  18& $\mathrm{4s^24p4d~^3F^o_4}$     & 125642.66 & 24.27  & 10.15 & 5.68  & $-$1.13 & $-$0.54  \\
  19& $\mathrm{4s^24p5s~^3P^o_2}$     & 128936.75 & $-$3.49  & 2.10  & 0.94  & $-$1.81 & $-$0.45  \\
  20& $\mathrm{4s^24p5s~^1P^o_1}$     & 130591.07 & $-$8.54  & $-$1.33 & $-$1.25 & $-$2.15 & $-$0.38  \\
  21& $\mathrm{4s^24p4d~^1P^o_1}$     & 137565.08 & 13.92  & 6.19  & 4.15  & 1.53  & 3.00   \\
  22& $\mathrm{4s^24p4d~^3P^o_2}$     & 139151.62 & 26.54  & 6.58  & 2.91  & $-$0.94 & 0.38   \\
  23& $\mathrm{4s^24p4d~^3D^o_1}$     & 139543.04 & 8.59   & 5.13  & 2.81  & 0.45  & 1.55   \\
  24& $\mathrm{4s^24p4d~^3P^o_1}$     & 141784.77 & 6.64   & 5.01  & 3.11  & 1.05  & 2.69   \\
  25& $\mathrm{4s^24p4d~^3D^o_3}$     & 141823.00 & 12.16  & 3.49  & 2.44  & $-$0.03 & 1.29   \\
  26& $\mathrm{4s^24p4d~^3P^o_0}$     & 141908.40 & 38.94  & 23.58 & 17.88 & 12.21 & 10.17  \\
  27& $\mathrm{4s^24p4d~^3D^o_2}$     & 142329.40 & 15.47  & 5.73  & 4.33  & 1.29  & 2.82   \\
  28& $\mathrm{4s4p^3~^3S^o_1}$       & 142728.82 & 10.36  & 6.44  & 5.07  & $-$0.09 & 1.11   \\
  29& $\mathrm{4s^24p4d~^1F^o_3}$     & 148254.58 & 33.45  & 10.95 & 5.85  & 2.92  & 3.60   \\
  30& $\mathrm{4s^24p5p~^1P_1}$       & 149010.74 & $-$8.20  & 1.30  & 0.89  & $-$0.28 & $-$0.47  \\
  31& $\mathrm{4s4p^3~^1D^o_2}$       & 149337.81 & 17.88  & 6.20  & 4.28  & 0.90  & 2.54   \\
  32& $\mathrm{4s^24p5p~^3D_1}$       & 151444.00 & $-$9.52  & 2.16  & 0.40  & $-$0.86 & $-$0.49  \\
  33& $\mathrm{4s^24p5p~^3D_2}$       & 151786.35 & $-$12.39 & 0.23  & $-$0.46 & $-$1.82 & $-$0.46  \\
  34& $\mathrm{4s^24p5p~^3P_0}$       & 153156.20 & $-$12.31 & $-$1.03 & $-$0.31 & $-$0.87 & $-$0.47  \\
  35& $\mathrm{4s^24p5p~^3P_1}$       & 154556.08 & $-$13.91 & 0.07  & $-$0.40 & $-$1.22 & $-$0.52  \\
  36& $\mathrm{4s^24p5p~^3D_3}$       & 154792.05 & 4.93   & 6.75  & 2.93  & $-$0.11 & $-$0.56  \\
  37& $\mathrm{4s^24p5p~^3P_2}$       & 156058.57 & $-$16.77 & $-$1.48 & $-$1.36 & $-$1.92 & $-$0.48  \\
  38& $\mathrm{4s^24p5p~^3S_1}$       & 157478.90 & $-$0.55  & 6.69  & 3.82  & 0.94  & $-$0.56  \\
  39& $\mathrm{4s4p^3~^1P^o_1}$       & 158678.10 & 26.45  & 8.49  & 5.33  & 0.72  & 1.88   \\
  40& $\mathrm{4s^24p5p~^1D_2}$       & 159471.51 & $-$13.02 & $-$0.67 & $-$0.87 & $-$1.83 & $-$0.34  \\
  41& $\mathrm{4s^24p5p~^1S_0}$       & 165027.72 & $-$1.67  & 2.69  & 0.97  & $-$0.72 & $-$0.49  \\
  42& $\mathrm{4s^24p6s~^3P^o_0}$     & 185422.73 & $-$0.79  & 6.18  & 5.72  & 3.98  & 4.50   \\
  43& $\mathrm{4s^24p6s~^3P^o_1}$     & 185652.50 & $-$13.70 & $-$1.92 & $-$1.80 & $-$1.88 & 0.02   \\
  44& $\mathrm{4s^24p5d~^3F^o_2}$     & 186497.20 & $-$9.91  & 0.89  & 1.01  & $-$0.64 & 0.89   \\
  45& $\mathrm{4s^24p5d~^3F^o_3}$     & 187756.01 & $-$11.57 & $-$0.81 & $-$1.29 & $-$1.78 & $-$0.19  \\
  46& $\mathrm{4s^24p4f~^1F_3}$       & 187772.82 & 2.24   & 6.19  & 2.96  & 0.07  & 0.84   \\
  47& $\mathrm{4s^24p4f~^3F_3}$       & 188094.45 & 1.68   & 1.21  & 1.20  & $-$1.14 & $-$0.24  \\
  48& $\mathrm{4s^24p4f~^3F_2}$       & 188261.84 & $-$4.91  & 0.21  & 0.90  & $-$0.42 & 1.41   \\
  49& $\mathrm{4s^24p5d~^3D_2}$       & 188357.90 & $-$12.60 & $-$2.21 & $-$1.68 & $-$2.05 & $-$0.26  \\
  50& $\mathrm{4s^24p4f~^3F_4}$       & 188358.89 & $-$2.94  & 3.31  & 0.74  & $-$1.71 & $-$0.52  \\
  51& $\mathrm{4s^24p5d~^3D^o_1}$     & 189181.95 & $-$12.25 & $-$2.03 & $-$1.26 & $-$1.11 & 0.64   \\
  52& $\mathrm{4s^24p6s~^3P^o_2}$     & 189632.21 & $-$7.87  & 0.09  & 0.06  & $-$1.81 & $-$0.34  \\
  53& $\mathrm{4s4p^2~^4P~4d~^3P_2}$  & 189825.92 & $-$17.91 & $-$1.62 & $-$2.00 & $-$1.62 & 0.20   \\
  54& $\mathrm{4s^24p6s~^1P^o_1}$     & 190303.32 & $-$14.79 & $-$2.77 & $-$2.18 & $-$2.05 & $-$0.21  \\
  55& $\mathrm{4s^24p5d~^3F^o_4}$     & 190692.95 & $-$6.48  & 0.69  & $-$0.09 & $-$1.92 & $-$0.57  \\
  56& $\mathrm{4s4p^2~^4P~4d~^5F_1}$  & 191122.74 & $-$12.03 & 0.36  & 0.23  & $-$0.76 & 0.52   \\
  57& $\mathrm{4s^24p4f~^3G_3}$       & 191415.95 & 2.29   & 5.36  & 3.66  & 0.85  & 1.55   \\
  58& $\mathrm{4s^24p5d~^1D^o_2}$     & 191470.19 & $-$13.06 & $-$1.38 & $-$0.87 & $-$1.24 & 0.50   \\
  59& $\mathrm{4s4p^2~^4P~4d~^5F_2}$  & 191604.93 & $-$19.08 & $-$1.68 & $-$1.40 & $-$1.55 & 0.34   \\
  60& $\mathrm{4s^24p4f~^3G_4}$       & 191736.11 & $-$14.11 & 1.67  & 1.57  & $-$1.47 & $-$0.37  \\
  61& $\mathrm{4s4p^2~^4P~4d~^3P_1}$  & 191738.83 & $-$3.89  & 1.40  & $-$0.89 & $-$1.54 & $-$0.04  \\
  62& $\mathrm{4s^24p5d~^3D^o_3}$     & 192143.31 & $-$11.76 & $-$1.12 & $-$0.66 & $-$1.46 & 0.20   \\
  63& $\mathrm{4s4p^2~^4P~4d~^5F_3}$  & 192338.87 & $-$16.26 & $-$0.87 & $-$0.47 & $-$0.92 & 0.10   \\
  64& $\mathrm{4s4p^2~^4P~4d~^3P_0}$  & 192422.20 & $-$12.00 & $-$1.54 & $-$1.11 & $-$1.52 & $-$0.42  \\
  65& $\mathrm{4s^24p4f~^3G_5}$       & 192463.15 & 12.30  & 10.95 & 5.33  & $-$0.06 & $-$0.53  \\
  66& $\mathrm{4s^24p5d~^3P^o_2}$     & 192891.28 & $-$12.17 & $-$0.81 & $-$0.75 & $-$1.34 & 0.38   \\
  67& $\mathrm{4s^24p4f~^3D_3}$       & 193006.39 & 1.65   & 2.31  & 1.20  & $-$1.11 & $-$0.40  \\
  68& $\mathrm{4s^24p5d~^3P^o_1}$     & 193091.15 & $-$8.90  & 2.03  & 1.11  & $-$0.55 & 1.02   \\
  69& $\mathrm{4s^24p4f~^3D_2}$       & 193261.09 & $-$2.49  & 1.34  & 0.98  & $-$0.70 & 0.94   \\
  70& $\mathrm{4s^24p5d~^3P^o_0}$     & 193289.48 & $-$0.40  & 4.21  & 3.18  & 1.84  & 2.26   \\
  71& $\mathrm{4s4p^2~^4P~4d~^5F _4}$ & 193383.00 & $-$17.47 & $-$1.98 & $-$1.29 & $-$1.73 & $-$0.23  \\
  72& $\mathrm{4s^24p4f~^1G_4}$       & 193831.70 & 3.57   & 6.94  & 4.42  & $-$1.02 & $-$0.11  \\
  73& $\mathrm{4s^24p4f~^3D_1}$       & 194065.96 & 3.71   & 2.07  & 1.79  & $-$0.78 & 0.38   \\
  74& $\mathrm{4s^24p4f~^1D_2}$       & 194417.51 & $-$6.59  & 1.25  & 0.28  & $-$2.13 & $-$0.46  \\
  75& $\mathrm{4s4p^2~^4P~4d~^5F_5}$  & 194813.38 & $-$7.85  & 3.22  & 2.41  & 0.15  & $-$0.47  \\
  76& $\mathrm{4s^24p5d~^1F^o_3}$     & 195308.85 & 5.19   & 3.45  & 1.61  & 0.14  & 1.27   \\
  77& $\mathrm{4s^24p6p~^3D_1}$       & 195326.09 & $-$1.39  & 3.10  & 1.70  & $-$0.45 & $-$0.32  \\
  78& $\mathrm{4s^24p6p~^3P_1}$       & 196089.89 & $-$7.82  & 1.26  & 0.16  & $-$1.46 & $-$0.46  \\
  79& $\mathrm{4s^24p6p~^3D_2}$       & 196363.01 & $-$12.02 & $-$0.72 & $-$0.83 & $-$2.19 & $-$0.46  \\
  80& $\mathrm{4s^24p5d~^1P^o_1}$     & 196449.92 & $-$0.06  & 2.23  & 1.97  & $-$0.32 & 1.09   \\
  81& $\mathrm{4s^24p6p~^3P_0}$        & 196487.56 & $-$11.57 & $-$0.66 & 0.15  & $-$1.21 & $-$0.49  \\
  82& $\mathrm{4s4p^2~^4P~4d~^5D_0}$  & 197575.86 & $-$12.60 & 0.14  & 0.20  & $-$1.06 & 0.16   \\
  83& $\mathrm{4s4p^2~^4P~4d~^5D_1}$  & 197625.39 & $-$14.48 & $-$0.26 & $-$0.13 & $-$1.04 & 0.22   \\
  84& $\mathrm{4s4p^2~^4P~4d~^5D_2}$  & 197799.40 & $-$18.18 & $-$1.91 & $-$1.24 & $-$1.67 & 0.14   \\
  85& $\mathrm{4s4p^2~^4P~4d~^5D_3}$  & 198140.01 & $-$10.09 & 0.29  & 0.53  & $-$0.60 & $-$0.06  \\
  86& $\mathrm{4s4p^2~^4P~4d~^5D_4}$  & 198769.40 & $-$11.27 & $-$0.27 & 0.13  & $-$1.24 & $-$0.31  \\
  87& $\mathrm{4s^24p6p~^1P_1}$       & 199127.90 & $-$10.64 & 0.98  & 0.18  & $-$1.43 & $-$0.41  \\
  88& $\mathrm{4s4p^2~^4P~5s~^5P_1}$  & 199434.62 & $-$14.91 & 0.78  & 0.54  & $-$1.67 & $-$0.37  \\
  89& $\mathrm{4s^24p6p~^3P_2}$       & 199670.12 & $-$11.22 & 0.39  & $-$0.72 & $-$1.93 & $-$0.34  \\
  90& $\mathrm{4s^24p6p~^3D_3}$       & 199897.17 & 2.05   & 4.32  & 1.51  & $-$0.96 & $-$0.54  \\
  91& $\mathrm{4p^4~^1D_2}$           & 200121.14 & $-$3.80  & 8.54  & 3.98  & $-$1.41 & 0.08   \\
  92& $\mathrm{4s^24p6p~^3S_1}$       & 200722.94 & $-$7.85  & 1.67  & 0.84  & $-$0.65 & $-$0.55  \\
  93& $\mathrm{4s4p^2~^4P~5s~^5P_2}$  & 200855.25 & $-$24.47 & $-$2.05 & $-$0.92 & $-$2.23 & $-$0.33  \\
  94& $\mathrm{4s^24p6p~^1D_2}$       & 201357.48 & $-$11.77 & $-$0.23 & $-$0.85 & $-$2.04 & $-$0.40  \\
  95& $\mathrm{4s4p^2~^4P~5s~^5P_3}$  & 202670.04 & $-$20.50 & $-$1.74 & $-$0.88 & $-$1.59 & $-$0.42  \\
  96& $\mathrm{4s^24p6p~^1S_0}$       & 203344.68 & $-$1.81  & 2.48  & 1.99  & $-$0.75 & $-$0.43  \\
  97& $\mathrm{4s4p^2~^4P~4d~^3F_2}$  & 203981.62 & 5.51   & 18.23 & 15.65 & $-$0.42 & 0.60   \\
  98& $\mathrm{4s4p^2~^4P~4d~^3F_3}$  & 205082.24 & 20.97  & 16.09 & 15.70 & 4.22  & 0.44   \\
  99& $\mathrm{4s4p^2~^4P~4d~^3F_4}$  & 206668.27 & 18.28  & 11.00 & 6.38  & $-$0.20 & 0.18   \\
 100& $\mathrm{4s4p^2~^4P~5s~^3P_0}$  & 207298.91 & 2.94   & 6.84  & 7.02  & 0.99  & $-$0.04  \\
 101& $\mathrm{4s4p^2~^4P~5s~^3P_1}$  & 208435.16 & 5.54   & 12.99 & 8.55  & $-$1.06 & $-$0.27  \\
 102& $\mathrm{4s4p^2~^4P~4d~^5P_3}$  & 208999.03 & $-$14.23 & $-$1.19 & $-$0.42 & $-$0.99 & $-$0.29  \\
 103& $\mathrm{4s4p^2~^4P~4d~^5P_2}$  & 209605.76 & $-$22.14 & $-$2.30 & $-$1.36 & $-$1.56 & 0.22   \\
 104& $\mathrm{4s4p^2~^4P~4d~^5P_1}$  & 210090.41 & $-$17.96 & $-$0.47 & 0.15  & $-$0.34 & 0.56   \\
 105& $\mathrm{4s4p^2~^4P~5s~^3P_2}$  & 210575.00 & $-$3.50  & 5.52  & 2.09  & $-$1.98 & $-$0.16  \\
\hline
\noalign{\smallskip}
\multicolumn{2}{l}{$N_{CSFs}$} &    285503&    176181&	219451&	232906&	259552& 274526  \\
\end{longtable}
}

\begin{table*}[!ht]
{\scriptsize
\caption{The total energies (in a.u.) from \textbf{CV+C+CC+VV~RCI} calculations and differences (in a.u.) between \textbf{CV+C+CC+VV~RCI~(RSMBPT)} and \textbf{CV+C+CC+VV~RCI} 
 energies ($\Delta E_{\textbf{(CV+C+CC+VV~RCI~(RSMBPT))-(CV+C+CC+VV~RCI)}}$) for Se~III are given when CV, C, CC and VV correlations are included in the computations.}            
\label{comp3}
\centering
\begin{tabular}{r l r r r r r r}
\hline\hline
\multicolumn{1}{c}{\multirow{2}{*}{No.}} & \multicolumn{1}{c}{\multirow{2}{*}{State}} & \multicolumn{1}{c}{\multirow{2}{*}{\textbf{CV+C+CC+VV~RCI}}}& \multicolumn{5}{c}{$\Delta E_{\textbf{(CV+C+CC+VV~RCI~(RSMBPT))-(CV+C+CC+VV~RCI)}}$} \\
\cline{4-8}
&&& \multicolumn{1}{c}{95\%} & \multicolumn{1}{c}{99\%} & \multicolumn{1}{c}{99.5\%} & \multicolumn{1}{c}{99.95\%} & \multicolumn{1}{c}{100\%} \\
\hline
\noalign{\smallskip}
  1& $\mathrm{4s^24p^2~^3P_0}$       & -2426.0472568 & 0.0189950 & 0.0037583 & 0.0019239 & 0.0002134 & 0.0000042  \\
  2& $\mathrm{4s^24p^2~^1S_0}$       & -2425.9140952 & 0.0179700 & 0.0034686 & 0.0017785 & 0.0002033 & 0.0000018  \\
  3& $\mathrm{4s4p^3~^3D^o_1}$       & -2425.5492109 & 0.0149259 & 0.0029414 & 0.0014873 & 0.0001746 & 0.0000100  \\
  4& $\mathrm{4s4p^3~^3P^o_1}$       & -2425.4833650 & 0.0146271 & 0.0028168 & 0.0014096 & 0.0001598 & 0.0000076  \\
  5& $\mathrm{4s^24p5s~^3P^o_1}$     & -2425.3866923 & 0.0144394 & 0.0026780 & 0.0013029 & 0.0001423 & 0.0000140  \\
  6& $\mathrm{4s^24p5s~^1P^o_1}$     & -2425.3632499 & 0.0141322 & 0.0025139 & 0.0012108 & 0.0001314 & 0.0000092  \\
  7& $\mathrm{4s^24p5p~^3P_0}$       & -2425.3379284 & 0.0189785 & 0.0036897 & 0.0018012 & 0.0001797 & 0.0000040  \\
  8& $\mathrm{4s4p^3~^1P^o_1}$       & -2425.3326795 & 0.0145635 & 0.0028397 & 0.0014578 & 0.0001992 & 0.0000183  \\
  9& $\mathrm{4s^24p4d~^3D^o_1}$     & -2425.3215444 & 0.0145007 & 0.0028245 & 0.0014528 & 0.0002087 & 0.0000189  \\
 10& $\mathrm{4s4p^3~^3S^o_1}$       & -2425.3136899 & 0.0142691 & 0.0027807 & 0.0014103 & 0.0001843 & 0.0000141  \\
 11& $\mathrm{4s^24p4d~^3P^o_1}$     & -2425.3074798 & 0.0143708 & 0.0027160 & 0.0013700 & 0.0001977 & 0.0000263  \\
 12& $\mathrm{4s^24p5p~^1S_0}$       & -2425.2811712 & 0.0188732 & 0.0036309 & 0.0017519 & 0.0001657 & 0.0000012  \\
 13& $\mathrm{4s^24p4d~^1P^o_1}$     & -2425.2300584 & 0.0145587 & 0.0028452 & 0.0014526 & 0.0001891 & 0.0000183  \\
 14& $\mathrm{4s4p^2~^4P~4d~^3P_0}$  & -2425.1565596 & 0.0183315 & 0.0035864 & 0.0017834 & 0.0001845 & 0.0000022  \\
 15& $\mathrm{4s^24p6p~^3P_0}$       & -2425.1392487 & 0.0194013 & 0.0039494 & 0.0019745 & 0.0002050 & 0.0000045  \\
 16& $\mathrm{4s4p^2~^4P~4d~^5D_0}$  & -2425.1336409 & 0.0174931 & 0.0032964 & 0.0016133 & 0.0001691 & 0.0000054  \\
 17& $\mathrm{4s^24p6s~^3P^o_1}$     & -2425.1128679 & 0.0147355 & 0.0028337 & 0.0013878 & 0.0001399 & 0.0000142  \\
 18& $\mathrm{4s^24p6p~^1S_0}$       & -2425.1066908 & 0.0193385 & 0.0039076 & 0.0019286 & 0.0001819 & 0.0000011  \\
 19& $\mathrm{4s^24p5d~^3D^o_1}$     & -2425.0959911 & 0.0146767 & 0.0029321 & 0.0014872 & 0.0001649 & 0.0000112  \\
 20& $\mathrm{4s^24p6s~^1P^o_1}$     & -2425.0909697 & 0.0144467 & 0.0027629 & 0.0013389 & 0.0001358 & 0.0000095  \\
 21& $\mathrm{4s4p^2~^4P~5s~^3P_0}$  & -2425.0869795 & 0.0180387 & 0.0034797 & 0.0017028 & 0.0001766 & 0.0000033  \\
 22& $\mathrm{4s^24p5d~^3P^o_1}$     & -2425.0777530 & 0.0144899 & 0.0028563 & 0.0014218 & 0.0001554 & 0.0000105  \\
 23& $\mathrm{4s^24p5d~^1P^o_1}$     & -2425.0595874 & 0.0148405 & 0.0029726 & 0.0014743 & 0.0001577 & 0.0000099  \\
\hline
\noalign{\smallskip}
\multicolumn{2}{l}{$N_{CSFs}$} &    4456890&    1011352& 2055300&  2438634& 3317890&  4341267  \\
\hline
\hline
\end{tabular}
}
\end{table*}

\begin{table*}[!ht]
{\scriptsize
\caption{The energy levels (in cm$^{-1}$) from \textbf{CV+C+CC+VV~RCI} calculations and differences (in cm$^{-1}$) between \textbf{CV+C+CC+VV~RCI~(RSMBPT)} and \textbf{CV+C+CC+VV~RCI} 
 energies ($\Delta E_{\textbf{(CV+C+CC+VV~RCI~(RSMBPT))-(CV+C+CC+VV~RCI)}}$) for Se~III are given when CV, C, CC and VV correlations are included in the computations.}            
\label{comp4}
\centering
\begin{tabular}{r l r r r r r r}
\hline\hline
\multicolumn{1}{c}{\multirow{2}{*}{No.}} & \multicolumn{1}{c}{\multirow{2}{*}{State}} & \multicolumn{1}{c}{\multirow{2}{*}{\textbf{CV+C+CC+VV~RCI}}}& \multicolumn{5}{c}{$\Delta E_{\textbf{(CV+C+CC+VV~RCI~(RSMBPT))-(CV+C+CC+VV~RCI)}}$} \\
\cline{4-8}
&&& \multicolumn{1}{c}{95\%} & \multicolumn{1}{c}{99\%} & \multicolumn{1}{c}{99.5\%} & \multicolumn{1}{c}{99.95\%} & \multicolumn{1}{c}{100\%} \\
\hline
\noalign{\smallskip}
  1& $\mathrm{4s^24p^2~^3P_0}$       & 0         & 0        & 0       & 0       & 0      & 0      \\
  2& $\mathrm{4s^24p^2~^1S_0}$       & 29225.6   & $-$224.97  & $-$63.59  & $-$31.94  & $-$2.23  & $-$0.53  \\
  3& $\mathrm{4s4p^3~^3D^o_1}$       & 109308.44 & $-$893.06  & $-$179.27 & $-$95.84  & $-$8.52  & 1.27   \\
  4& $\mathrm{4s4p^3~^3P^o_1}$       & 123759.95 & $-$958.66  & $-$206.64 & $-$112.88 & $-$11.78 & 0.75   \\
  5& $\mathrm{4s^24p5s~^3P^o_1}$     & 144977.15 & $-$999.85  & $-$237.09 & $-$136.3  & $-$15.61 & 2.17   \\
  6& $\mathrm{4s^24p5s~^1P^o_1}$     & 150122.16 & $-$1067.25 & $-$273.1  & $-$156.51 & $-$18    & 1.12   \\
  7& $\mathrm{4s^24p5p~^3P_0}$       & 155679.59 & $-$3.63    & $-$15.05  & $-$26.93  & $-$7.39  & $-$0.04  \\
  8& $\mathrm{4s4p^3~^1P^o_1}$       & 156831.60 & $-$972.62  & $-$201.61 & $-$102.32 & $-$3.13  & 3.1    \\
  9& $\mathrm{4s^24p4d~^3D^o_1}$     & 159275.46 & $-$986.39  & $-$204.95 & $-$103.41 & $-$1.03  & 3.24   \\
 10& $\mathrm{4s4p^3~^3S^o_1}$       & 160999.33 & $-$1037.22 & $-$214.55 & $-$112.73 & $-$6.39  & 2.18   \\
 11& $\mathrm{4s^24p4d~^3P^o_1}$     & 162362.30 & $-$1014.91 & $-$228.77 & $-$121.59 & $-$3.47  & 4.84   \\
 12& $\mathrm{4s^24p5p~^1S_0}$       & 168136.36 & $-$26.74   & $-$27.97  & $-$37.76  & $-$10.48 & $-$0.65  \\
 13& $\mathrm{4s^24p4d~^1P^o_1}$     & 179354.33 & $-$973.66  & $-$200.4  & $-$103.45 & $-$5.34  & 3.09   \\
 14& $\mathrm{4s4p^2~^4P~4d~^3P_0}$  & 195485.44 & $-$145.64  & $-$37.73  & $-$30.84  & $-$6.36  & $-$0.42  \\
 15& $\mathrm{4s^24p6p~^3P_0}$       & 199284.75 & 89.16    & 41.94   & 11.09   & $-$1.86  & 0.06   \\
 16& $\mathrm{4s4p^2~^4P~4d~^5D_0}$  & 200515.51 & $-$329.62  & $-$101.36 & $-$68.18  & $-$9.73  & 0.28   \\
 17& $\mathrm{4s^24p6s~^3P^o_1}$     & 205074.67 & $-$934.87  & $-$202.94 & $-$117.67 & $-$16.16 & 2.19   \\
 18& $\mathrm{4s^24p6p~^1S_0}$       & 206430.38 & 75.38    & 32.76   & 1.02    & $-$6.93  & $-$0.66  \\
 19& $\mathrm{4s^24p5d~^3D^o_1}$     & 208778.70 & $-$947.76  & $-$181.34 & $-$95.86  & $-$10.66 & 1.53   \\
 20& $\mathrm{4s^24p6s~^1P^o_1}$     & 209880.76 & $-$998.23  & $-$218.46 & $-$128.41 & $-$17.04 & 1.17   \\
 21& $\mathrm{4s4p^2~^4P~5s~^3P_0}$  & 210756.50 & $-$209.89  & $-$61.13  & $-$48.53  & $-$8.09  & $-$0.19  \\
 22& $\mathrm{4s^24p5d~^3P^o_1}$     & 212781.48 & $-$988.76  & $-$197.95 & $-$110.2  & $-$12.73 & 1.39   \\
 23& $\mathrm{4s^24p5d~^1P^o_1}$     & 216768.39 & $-$911.83  & $-$172.45 & $-$98.69  & $-$12.23 & 1.25   \\
\hline
\noalign{\smallskip}
\multicolumn{2}{l}{$N_{CSFs}$} &    4456890&    1011352& 2055300&  2438634& 3317890&  4341267  \\
\multicolumn{2}{l}{rms (in cm$^{-1}$)} && 787.54&	170.67& 94.63&	10.16&  1.90  \\
\hline
\hline
\end{tabular}
}
\end{table*}

\begin{figure}[H]
\centering
\includegraphics[scale=0.4]{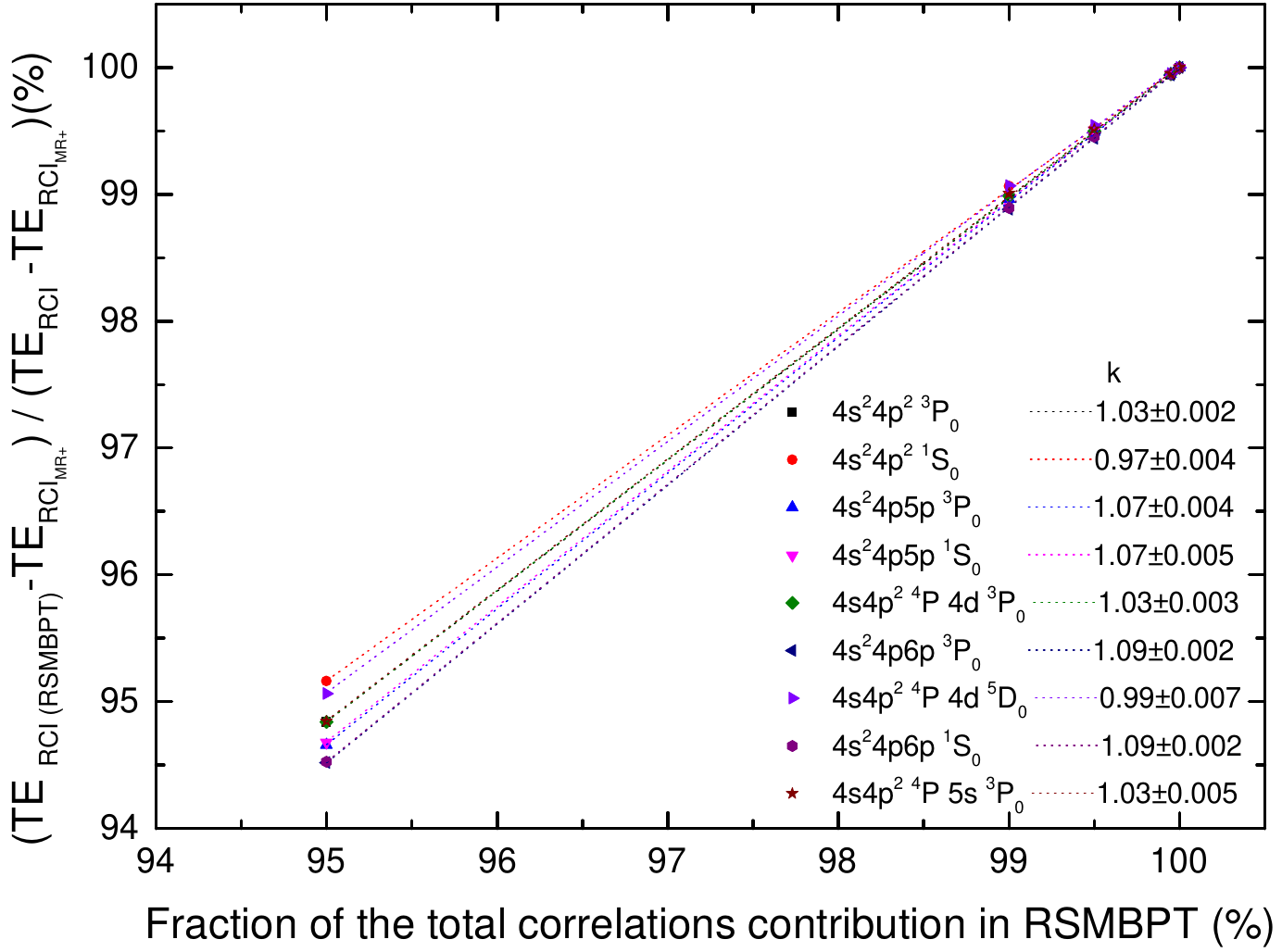}
\caption{\label{convergence_PT} The dependence of the contribution of the included correlations (in the case of CV+C+CC+VV) 
on the specified fraction of the total correlations contribution used in the \textbf{RCI~(RSMBPT)} method for levels of the even configurations.}  
\end{figure}

Figure \ref{convergence_PT} displays the contribution of the included correlations ($\Delta_{cor}$) using 
the \textbf{RCI~(RSMBPT)} method comparing to complete regular \textbf{RCI} computations, as a function of
the specified fraction of the total correlations contribution used in the \textbf{RCI~(RSMBPT)} method. 
This dependence is demonstrated for levels of the even configurations with $J$ = 0 in the case when the CV, C, CC and VV correlations are included.
The contribution $\Delta_{cor}$ (in \%) is computed according to the 
($TE_{\textbf{RCI~(RSMBPT)}}-TE_{\textbf{MR+}})/(TE_{\textbf{RCI}}-TE_{\textbf{MR+}})$, where
TE represents the total energy from a certain computational scheme. 
$TE_{\textbf{RCI~(RSMBPT)}}$ is the total energy from the \textbf{RCI~(RSMBPT)} method 
using the specified fraction (in \%) of the total correlations contribution.
The results of the \textbf{MR+} are from the computations when the CSFs basis consists 
of the MR set together with the correlations which were not included using the RSMBPT method. 
As can be observed in the figure, the computed $\Delta_{cor}$ is almost identical to the specified fraction (in \%) 
of the total correlations contribution used in the RCI (RSMBPT) method, exhibiting a linear dependence. 
The linear fitting ($y=kx+a$) demonstrates that the slope coefficient ($k$) is approaching unity 
and that the error associated with the fitting itself 
is negligible in comparison to the resulting value (see Fig. \ref{convergence_PT}), 
indicating that the $k$-factor is highly precise.
Based on the above regularities, it can be said 
that by choosing the preferred amount of the correlations using the RSMBPT method, 
the error of the total energy (the difference of TE computed by incorporating 100\% of the correlations
and reduced amount of the correlations) arising from the CSF basis reduction can be determined. 
Moreover, these regularities are stable for configurations with different $n$, $l$, and shells occupation.

Table \ref{convergence2} presents the total energies from \textbf{MR+} 
and regular \textbf{RCI} computations, when CV, C, CC and VV are included (\textbf{CV+C+CC+VV~RCI} strategy). 
The table also gives the calculated $\Delta_{cor}$ (in \%). 
All these results are given for levels of the even configurations 
with $J$ = 0 and for levels of the odd configurations with $J$ = 1.
As can be seen from the Table, the computed contribution of the included correlations ($\Delta_{cor}$ (in \%)) corresponds 
to the value which is used in the \textbf{RCI~(RSMBPT)} method for all computed levels.
By employing the linear dependence, it is feasible to ascertain 
not only the discrepancy in the total energy but also the value of the total energy itself. 
By taking results only from smaller bases that includes 95\% and 99\% of correlations using RSMBPT method,
it is possible to extrapolate the total energy value that would have been obtained if 100\% of the correlations would be included.
This could be very important for the study of energy levels that are close to each other.
These extrapolated values and the differences with the total energies computed 
by incorporating 100\% of the correlations using the RSMBPT method are given in Table \ref{convergence2} (last two columns, respectively).
The differences in the last column of the Table are small and reach up to 3.9E-04 a.u. ( or 0.000016\%).
One of the reasons of the disagreement could be the fact that while using the RSMBPT method
the impact of off-diagonal matrix elements is not taken into account.
Similar computations were performed without the Breit and QED effects, the trends of the results are the same.

Table \ref{comp_accum} presents the total energies 
from regular \textbf{CV+C+CC+VV~RCI}, differences between \textbf{CV+C+CC+VV~RCI~(RSMBPT)}, 
\textbf{CV+C+CC+VV~RCI~(accum)} with \textbf{CV+C+CC+VV~RCI} calculations. 
Results in columns '$\Delta E_{\textbf{(CV+C+CC+VV~RCI~(accum))-(CV+C+CC+VV~RCI)}}$' presents the results from 
'rmixaccumulate' program (which is included in the {\sc Grasp} package).
This program accumulates dominating CSFs by mixing coefficients up to
a user defined fraction of the atomic state function. In the Table few different values 
to accumulate CSF basis are taken as the example.
In the last lines of the table the number of CSFs, the rms, the largest ($\Delta E_{max}$) and 
the smallest ($\Delta E_{min}$) differences with the results of the regular 
{\sc Grasp}2018 calculations (\textbf{CV+C+CC+VV RCI}) are given for each computation. 
As seen from the Table the differences between \textbf{CV+C+CC+VV~RCI~(RSMBPT)} and \textbf{CV+C+CC+VV~RCI}
are similar for all levels and there is almost no scatter in the results with the same 
fraction of the total correlations contribution used in the RSMBPT method. For example, in the '95\%' case
the rms is equal to 0.0186129, $\Delta E_{max}$ -- 0.0194013 and $\Delta E_{min}$ -- 0.0174931.
Comparing \textbf{CV+C+CC+VV~RCI~(accum)} results with \textbf{CV+C+CC+VV~RCI}, 
the change in total energies for all levels is different in the same calculation using 'rmixaccumulate' program.
For example, in the '0.99947' case the rms is equal to 0.0100674, $\Delta E_{max}$ -- 0.0195031 and $\Delta E_{min}$ -- 0.0020293.
Finding the analogous calculation for \textbf{CV+C+CC+VV~RCI~(RSMBPT)} and \textbf{CV+C+CC+VV~RCI~(accum)} is
quite complex since using RSMBPT the selection process is based on configurations meanwhile 'rmixaccumulate' program
is based on CSFs. It should be mentioned, that in both computations only CSFs with non-zero matrix elements in the sets of spin-angular
integration with the CSFs belonging to the configurations in the MR are included. 
Thus, performing the \textbf{CV+C+CC+VV~RCI~(RSMBPT)} computations firstly the RSMBPT procedure, 
which is based on configurations, was applied 
and after the reductions of CSFs basis was done based on CSFs.
For example, two computations \textbf{CV+C+CC+VV~RCI~(RSMBPT)} and \textbf{CV+C+CC+VV~RCI~(accum)} have similar
rms deviation with the regular RCI computation 0.0036463 in the case of 99\% using RSMBPT, and 0.0036391 in the case of
'rmixaccumulate' program with a fraction equal to 0.99982.
The similar largest difference with regular RCI computation is 0.0194013 in case of 95\% using RSMBPT, and 0.0195031 in case of 
'rmixaccumulate' program with fraction equal to 0.99947. But min value and rms is very different.

\begin{table*}[!ht]
\setlength{\tabcolsep}{3pt}
{\scriptsize
\caption{The total energies (in a.u.) from \textbf{MR+} and regular \textbf{RCI} (when CV+C+CC+VV are included) computations are given. Column $\Delta_{cor}$ (in \%) presents the contribution of included correlations using the \textbf{RCI~(RSMBPT)} method comparing to regular \textbf{RCI} computations. The penultimate column in the table shows the extrapolated total energy, the last column shows the differences (in a.u.) between \textbf{CV+C+CC+VV~RCI~(RSMBPT)} in case of '100\%' and  the extrapolated value. }            
\label{convergence2}
\centering
\begin{tabular}{r l r r r r r r r r r r r}
\hline\hline
\multicolumn{1}{c}{\multirow{2}{*}{No.}}& \multicolumn{1}{c}{\multirow{2}{*}{State}} & \multicolumn{2}{c}{TE} && \multicolumn{5}{c}{$\Delta_{cor}$} & \multicolumn{1}{c}{TE} & \multicolumn{1}{c}{Diff} \\
\cline{3-4} \cline{6-10}
&& \multicolumn{1}{c}{MR+} & \multicolumn{1}{c}{\textbf{RCI}} && \multicolumn{1}{c}{95\%} &  \multicolumn{1}{c}{99\%} & \multicolumn{1}{c}{99.5\%} &  \multicolumn{1}{c}{99.95\%} & \multicolumn{1}{c}{100\%} &  \multicolumn{1}{c}{Extr. to 100\%} \\
\hline
\noalign{\smallskip}
1  & $\mathrm{4s^24p^2~^3P_0}$      & -2425.6790685 & -2426.0472568 && 94.841 & 98.979 & 99.477 & 99.942 & 99.999 & -2426.0473077 &  0.0000551 \\
2  & $\mathrm{4s^24p^2~^1S_0}$      & -2425.5428084 & -2425.9140952 && 95.160 & 99.066 & 99.521 & 99.945 & 100.000& -2425.9142520 &  0.0001585 \\
7  & $\mathrm{4s^24p5p~^3P_0}$      & -2424.9826929 & -2425.3379284 && 94.657 & 98.961 & 99.493 & 99.949 & 99.999 & -2425.3380609 &  0.0001365 \\
12 & $\mathrm{4s^24p5p~^1S_0}$      & -2424.9265608 & -2425.2811712 && 94.678 & 98.976 & 99.506 & 99.953 & 100.000& -2425.2813509 &  0.0001809 \\
14 & $\mathrm{4s4p^2~^4P~4d~^3P_0}$ & -2424.8013581 & -2425.1565596 && 94.839 & 98.990 & 99.498 & 99.948 & 99.999 & -2425.1566595 &  0.0001021 \\
15 & $\mathrm{4s^24p6p~^3P_0}$      & -2424.7852302 & -2425.1392487 && 94.520 & 98.884 & 99.442 & 99.942 & 99.999 & -2425.1391623 & -0.0000819 \\
16 & $\mathrm{4s4p^2~^4P~4d~^5D_0}$ & -2424.7793276 & -2425.1336409 && 95.063 & 99.070 & 99.545 & 99.952 & 99.998 & -2425.1338937 &  0.0002582 \\
18 & $\mathrm{4s^24p6p~^1S_0}$      & -2424.7533700 & -2425.1066908 && 94.527 & 98.894 & 99.454 & 99.949 & 100.000& -2425.1066409 & -0.0000488 \\
21 & $\mathrm{4s4p^2~^4P~5s~^3P_0}$ & -2424.7371032 & -2425.0869795 && 94.844 & 99.005 & 99.513 & 99.950 & 99.999 & -2425.0871396 &  0.0001633 \\
3  & $\mathrm{4s4p^3~^3D^o_1}$      & -2425.2697629 & -2425.5492109 && 94.659 & 98.947 & 99.468 & 99.938 & 99.996 & -2425.5492656 &  0.0000647 \\
4  & $\mathrm{4s4p^3~^3P^o_1}$      & -2425.2000921 & -2425.4833650 && 94.836 & 99.006 & 99.502 & 99.944 & 99.997 & -2425.4835008 &  0.0001434 \\
5  & $\mathrm{4s^24p5s~^3P^o_1}$    & -2425.1080919 & -2425.3866923 && 94.817 & 99.039 & 99.532 & 99.949 & 99.995 & -2425.3869547 &  0.0002764 \\
6  & $\mathrm{4s^24p5s~^1P^o_1}$    & -2425.0839277 & -2425.3632499 && 94.941 & 99.100 & 99.567 & 99.953 & 99.997 & -2425.3636406 &  0.0003999 \\
8  & $\mathrm{4s4p^3~^1P^o_1}$      & -2425.0483337 & -2425.3326795 && 94.878 & 99.001 & 99.487 & 99.930 & 99.994 & -2425.3327708 &  0.0001096 \\
9  & $\mathrm{4s^24p4d~^3D^o_1}$    & -2425.0429675 & -2425.3215444 && 94.795 & 98.986 & 99.478 & 99.925 & 99.993 & -2425.3216390 &  0.0001134 \\
10 & $\mathrm{4s4p^3~^3S^o_1}$      & -2425.0277709 & -2425.3136899 && 95.009 & 99.027 & 99.507 & 99.936 & 99.995 & -2425.3137813 &  0.0002757 \\
11 & $\mathrm{4s^24p4d~^3P^o_1}$    & -2425.0328347 & -2425.3074798 && 94.768 & 99.011 & 99.501 & 99.928 & 99.990 & -2425.3076775 &  0.0002240 \\
13 & $\mathrm{4s^24p4d~^1P^o_1}$    & -2424.9528259 & -2425.2300584 && 94.749 & 98.974 & 99.476 & 99.932 & 99.993 & -2425.2301416 &  0.0001015 \\
17 & $\mathrm{4s^24p6s~^3P^o_1}$    & -2424.8346996 & -2425.1128679 && 94.703 & 98.981 & 99.501 & 99.950 & 99.995 & -2425.1130097 &  0.0001560 \\
19 & $\mathrm{4s^24p5d~^3D^o_1}$    & -2424.8185077 & -2425.0959911 && 94.711 & 98.943 & 99.464 & 99.941 & 99.996 & -2425.0959952 &  0.0000152 \\
20 & $\mathrm{4s^24p6s~^1P^o_1}$    & -2424.8135056 & -2425.0909697 && 94.793 & 99.004 & 99.517 & 99.951 & 99.997 & -2425.0911278 &  0.0001675 \\
22 & $\mathrm{4s^24p5d~^3P^o_1}$    & -2424.8008383 & -2425.0777530 && 94.767 & 98.969 & 99.487 & 99.944 & 99.996 & -2425.0778051 &  0.0000626 \\
23 & $\mathrm{4s^24p5d~^1P^o_1}$    & -2424.7839472 & -2425.0595874 && 94.616 & 98.922 & 99.465 & 99.943 & 99.996 & -2425.0595818 &  0.0000043 \\
\hline
\hline
\end{tabular}
}
\end{table*}

\begin{table*}[!ht]
{\scriptsize
\caption{The total energies (in a.u.) from \textbf{CV+C+CC+VV~RCI} calculations and differences (in a.u.) 
between \textbf{CV+C+CC+VV~RCI~(RSMBPT)} and \textbf{CV+C+CC+VV~RCI} 
energies ($\Delta E_{\textbf{(CV+C+CC+VV~RCI~(RSMBPT))-(CV+C+CC+VV~RCI)}}$) and
between \textbf{CV+C+CC+VV~RCI~(accum)} and \textbf{CV+C+CC+VV~RCI} 
energies ($\Delta E_{\textbf{(CV+C+CC+VV~RCI~(accum))-(CV+C+CC+VV~RCI)}}$) 
for Se~III are given when CV, C, CC and VV correlations are included in the computations.}           
\label{comp_accum}
\centering
\begin{tabular}{r l r r r r r r}
\hline\hline
\multicolumn{1}{c}{\multirow{2}{*}{No.}} & \multicolumn{1}{c}{\multirow{2}{*}{State}} & \multicolumn{1}{c}{\multirow{2}{*}{\textbf{CV+C+CC+VV~RCI}}}& \multicolumn{5}{c}{$\Delta E_{\textbf{(CV+C+CC+VV~RCI~(RSMBPT))-(CV+C+CC+VV~RCI)}}$} \\
\cline{4-8}
&&& \multicolumn{1}{c}{95\%} & \multicolumn{1}{c}{99\%} & \multicolumn{1}{c}{99.5\%} & \multicolumn{1}{c}{99.95\%} & \multicolumn{1}{c}{100\%} \\
\hline
\noalign{\smallskip}
  1& $\mathrm{4s^24p^2~^3P_0}$       & -2426.0472568 & 0.0189950 & 0.0037583 & 0.0019239 & 0.0002134 & 0.0000042  \\
  2& $\mathrm{4s^24p^2~^1S_0}$       & -2425.9140952 & 0.0179700 & 0.0034686 & 0.0017785 & 0.0002033 & 0.0000018  \\
  3& $\mathrm{4s^24p5p~^3P_0}$       & -2425.3379284 & 0.0189785 & 0.0036897 & 0.0018012 & 0.0001797 & 0.0000040  \\
  4& $\mathrm{4s^24p5p~^1S_0}$       & -2425.2811712 & 0.0188732 & 0.0036309 & 0.0017519 & 0.0001657 & 0.0000012  \\
  5& $\mathrm{4s4p^2~^4P~4d~^3P_0}$  & -2425.1565596 & 0.0183315 & 0.0035864 & 0.0017834 & 0.0001845 & 0.0000022  \\
  6& $\mathrm{4s^24p6p~^3P_0}$       & -2425.1392487 & 0.0194013 & 0.0039494 & 0.0019745 & 0.0002050 & 0.0000045  \\
  7& $\mathrm{4s4p^2~^4P~4d~^5D_0}$  & -2425.1336409 & 0.0174931 & 0.0032964 & 0.0016133 & 0.0001691 & 0.0000054  \\
  8& $\mathrm{4s^24p6p~^1S_0}$       & -2425.1066908 & 0.0193385 & 0.0039076 & 0.0019286 & 0.0001819 & 0.0000011  \\
  9& $\mathrm{4s4p^2~^4P~5s~^3P_0}$  & -2425.0869795 & 0.0180387 & 0.0034797 & 0.0017028 & 0.0001766 & 0.0000033  \\
\hline
\noalign{\smallskip}
\multicolumn{2}{l}{$N_{CSFs}$} &    1303883&    298310& 598587&  711815& 978504&  1275596  \\
\multicolumn{2}{l}{rms (in a.u.)}    && 0.0186129&	0.0036463& 0.0018098&	0.0001872&  0.0000034  \\
\multicolumn{2}{l}{$\Delta E_{max}$ (in a.u.)} && 0.0194013&	0.0039494& 0.0019745&	0.0002134&  0.0000054  \\
\multicolumn{2}{l}{$\Delta E_{min}$ (in a.u.)} && 0.0174931&	0.0032964& 0.0016133&	0.0001657&  0.0000011  \\
\hline
\hline
\multicolumn{1}{c}{\multirow{2}{*}{No.}} & \multicolumn{1}{c}{\multirow{2}{*}{State}} & \multicolumn{1}{c}{\multirow{2}{*}{\textbf{CV+C+CC+VV~RCI}}}& \multicolumn{5}{c}{$\Delta E_{\textbf{(CV+C+CC+VV~RCI~(accum))-(CV+C+CC+VV~RCI)}}$} \\
\cline{4-8}
&&& \multicolumn{1}{c}{0.99947} & \multicolumn{1}{c}{0.9995} & \multicolumn{1}{c}{0.9997} & \multicolumn{1}{c}{0.99982} & \multicolumn{1}{c}{0.99996} \\
\hline
\noalign{\smallskip}
  1& $\mathrm{4s^24p^2~^3P_0}$       & -2426.0472568 & 0.0023977 & 0.0022523 & 0.0012670 & 0.0007436 & 0.0001736  \\             
  2& $\mathrm{4s^24p^2~^1S_0}$       & -2425.9140952 & 0.0020293 & 0.0018963 & 0.0009976 & 0.0005485 & 0.0001113  \\             
  3& $\mathrm{4s^24p5p~^3P_0}$       & -2425.3379284 & 0.0038803 & 0.0036183 & 0.0019734 & 0.0010759 & 0.0002032  \\             
  4& $\mathrm{4s^24p5p~^1S_0}$       & -2425.2811712 & 0.0054104 & 0.0050714 & 0.0028190 & 0.0015648 & 0.0002892  \\             
  5& $\mathrm{4s4p^2~^4P~4d~^3P_0}$  & -2425.1565596 & 0.0152710 & 0.0145833 & 0.0094772 & 0.0059936 & 0.0014514  \\             
  6& $\mathrm{4s^24p6p~^3P_0}$       & -2425.1392487 & 0.0064461 & 0.0059563 & 0.0031491 & 0.0017466 & 0.0003503  \\             
  7& $\mathrm{4s4p^2~^4P~4d~^5D_0}$  & -2425.1336409 & 0.0195031 & 0.0184791 & 0.0116459 & 0.0072768 & 0.0017522  \\             
  8& $\mathrm{4s^24p6p~^1S_0}$       & -2425.1066908 & 0.0067823 & 0.0063778 & 0.0037588 & 0.0021983 & 0.0004530  \\             
  9& $\mathrm{4s4p^2~^4P~5s~^3P_0}$  & -2425.0869795 & 0.0125242 & 0.0117847 & 0.0071323 & 0.0042394 & 0.0008940  \\             
\hline                                                                                                                                    
\noalign{\smallskip}                                                                                                             
\multicolumn{2}{l}{$N_{CSFs}$} &    1303883& 211374&    217926& 278203&  342510 &  539460  \\                                    
\multicolumn{2}{l}{rms (in a.u.)}              && 0.0100674&	0.0095258& 0.0059142&	0.0036391&  0.0008480  \\                     
\multicolumn{2}{l}{$\Delta E_{max}$ (in a.u.)} && 0.0195031&	0.0184791& 0.0116459&	0.0072768&  0.0017522  \\                     
\multicolumn{2}{l}{$\Delta E_{min}$ (in a.u.)} && 0.0020293&	0.0018963& 0.0009976&	0.0005485&  0.0001113  \\
\hline
\hline
\end{tabular}
}
\end{table*}

The transition properties of E1 transitions between the levels of the even configurations 
with $J$ = 0 and levels of the odd configurations with $J$ = 1 are also computed.
The calculations are done in a regular way without and with accumulate option and using 
the RSMBPT method when CV, C, CC and VV are included. 
Table \ref{com_line_st_schemes} shows the comparison of the line strengths, cancellation factors (CF) \cite{Cowan}, 
the $G_{S=0}$ parameters, and the estimated accuracy for few E1 transitions using different computational schemes.
The uncertainties of the line strengths obtained in this work are estimated based on the quantitative and qualitative evaluation
(QQE) method described in Refs. \cite{Ce_IV,Pr_IV,Se_Ge_like}.
By comparing the line strengths from the regular {\sc Grasp}2018 with results from 
\textbf{CV+C+CC+VV~RCI~(RSMBPT)} calculations in case of '100\%' it is seen that results using 
RSMBPT are reproduced. Excluding the CV, C, CC, VV correlations with the smallest impact (cases 99.95, 99.5 and 99 in Table \ref{com_line_st_schemes}), 
we see that the line strengths almost does not change comparing to the results when all these correlations are included.
From Table \ref{com_line_st_schemes} we also see that the CFs in both (Babushkin and Coulomb) gauges 
are stable when the most important configurations 
of CV, C, CC and VV correlations in different amount are included.
Comparing the line strengths from the regular {\sc Grasp}2018 with results from 
\textbf{CV+C+CC+VV~RCI~(accum)} calculations it is seen that results agree among themselves 
but differences with \textbf{CV+C+CC+VV~RCI} are larger than 
obtained with \textbf{CV+C+CC+VV~RCI~(RSMBPT)}.

{\scriptsize
\begin{longtable}{l l r r r r r l}
\caption{\label{com_line_st_schemes} Comparison of computed wavelengths ($\lambda$ in \AA), line strengths ($S$ in a.u.), 
cancellation factors, and the $G_{S=0}$ parameters using different strategies. 
Subscript $B$ means the Babushkin gauge, $C$ -- the Coulomb gauge. 
The name of the computational scheme \textbf{CV+C+CC+VV~RCI~(accum)} is marked as \textbf{RCI~(accum)}, 
the \textbf{CV+C+CC+VV~RCI(RSMBPT)} is marked as \textbf{RCI~(RSMBPT)} in the table.}\\
\hline\hline
\multicolumn{1}{c}{Strategy} & \multicolumn{1}{c}{$\lambda$} & \multicolumn{1}{c}{$S_B$} & \multicolumn{1}{c}{$S_C$} & \multicolumn{1}{c}{CF$_B$} & \multicolumn{1}{c}{CF$_C$} & \multicolumn{1}{c}{$G_{S=0}$} & \multicolumn{1}{c}{Acc.}\\
\hline
\noalign{\smallskip}
\endfirsthead
\caption{Continued.}\\
\hline\hline
\multicolumn{1}{c}{Strategy} & \multicolumn{1}{c}{$\lambda$} & \multicolumn{1}{c}{$S_B$} & \multicolumn{1}{c}{$S_C$} & \multicolumn{1}{c}{CF$_B$} & \multicolumn{1}{c}{CF$_C$} & \multicolumn{1}{c}{$G_{S=0}$} & \multicolumn{1}{c}{Acc.}\\
\hline
\noalign{\smallskip}
\endhead
\noalign{\smallskip}
\hline
\hline
\endfoot
\noalign{\smallskip}
\multicolumn{8}{c}{$\mathrm{4s4p^3~^3D^o_1}$ -- $\mathrm{4s4p^2~^4P~5s~^3P_0}$} \\
\textbf{CV+C+CC+VV~RCI}	          &985.73& 6.29208E-01& 1.00321E+00& 2.24E-01& 2.42E-01& 6.79773E+00& D+ \\
\textbf{RCI~(accum) 0.99996}   &985.37& 6.32634E-01& 1.00671E+00& 2.26E-01& 2.44E-01& 6.82291E+00& D+ \\
\textbf{RCI~(accum) 0.99982}   &983.59& 6.43904E-01& 1.01617E+00& 2.32E-01& 2.49E-01& 6.93329E+00& D+ \\
\textbf{RCI~(accum) 0.9997}   &981.89& 6.51795E-01& 1.01997E+00& 2.37E-01& 2.52E-01& 7.04982E+00& D+ \\
\textbf{RCI~(accum) 0.9995}   &979.31& 6.62218E-01& 1.02468E+00& 2.44E-01& 2.58E-01& 7.24111E+00& D+ \\
\textbf{RCI~(accum) 0.99947}   &978.85& 6.63737E-01& 1.02502E+00& 2.43E-01& 2.57E-01& 7.21203E+00& D+ \\
\textbf{RCI~(RSMBPT) 100\%}       &985.74& 6.29174E-01& 1.00326E+00& 2.23E-01& 2.42E-01& 6.79638E+00& D+ \\
\textbf{RCI~(RSMBPT) 99.95\%}	    &985.72& 6.29329E-01& 1.00362E+00& 2.24E-01& 2.42E-01& 6.79487E+00& D+ \\
\textbf{RCI~(RSMBPT) 99.5\%}      &985.27& 6.27428E-01& 1.00053E+00& 2.24E-01& 2.42E-01& 6.79565E+00& D+ \\
\textbf{RCI~(RSMBPT) 99\%}	      &984.58& 6.25883E-01& 9.97199E-01& 2.23E-01& 2.43E-01& 6.80689E+00& D+ \\
\textbf{RCI~(RSMBPT) 95\%}	      &979.13& 6.19460E-01& 9.74757E-01& 2.21E-01& 2.44E-01& 6.97288E+00& D+ \\
\multicolumn{8}{c}{$\mathrm{4s4p^3~^3D^o_1}$ -- $\mathrm{4s^24p6p~^3P_0}$} \\
\textbf{CV+C+CC+VV~RCI}	          &1111.40& 1.97065E-01& 2.18151E-01& 8.30E-02& 6.41E-02& 2.85371E+01& B \\
\textbf{RCI~(accum) 0.99996}   &1112.43& 2.00373E-01& 2.21194E-01& 8.35E-02& 6.39E-02& 2.93229E+01& B \\
\textbf{RCI~(accum) 0.99982}   &1115.45& 2.09708E-01& 2.30566E-01& 7.76E-02& 5.67E-02& 3.05418E+01& B \\
\textbf{RCI~(accum) 0.9997}   &1117.34& 2.18585E-01& 2.40375E-01& 6.97E-02& 4.84E-02& 3.04770E+01& B \\
\textbf{RCI~(accum) 0.9995}   &1119.05& 2.33846E-01& 2.58968E-01& 5.51E-02& 3.48E-02& 2.84319E+01& B \\
\textbf{RCI~(accum) 0.99947}   &1119.13& 2.35264E-01& 2.60959E-01& 5.29E-02& 3.48E-02& 2.84319E+01& B \\
\textbf{RCI~(RSMBPT) 100\%}       &1111.42& 1.97088E-01& 2.18166E-01& 8.30E-02& 6.41E-02& 2.85512E+01& B \\
\textbf{RCI~(RSMBPT) 99.95\%}	    &1111.32& 1.97034E-01& 2.18063E-01& 8.30E-02& 6.41E-02& 2.86052E+01& B \\
\textbf{RCI~(RSMBPT) 99.5\%}      &1110.08& 1.95504E-01& 2.16309E-01& 8.18E-02& 6.32E-02& 2.86824E+01& B \\
\textbf{RCI~(RSMBPT) 99\%}	      &1108.68& 1.93800E-01& 2.13872E-01& 8.05E-02& 6.22E-02& 2.94122E+01& B \\
\textbf{RCI~(RSMBPT) 95\%}	      &1099.40& 1.82939E-01& 1.99521E-01& 7.21E-02& 5.58E-02& 3.33109E+01& B \\
\multicolumn{8}{c}{$\mathrm{4s^24p5s~^3P^o_1}$ -- $\mathrm{4s^24p6p~^1S_0}$} \\
\textbf{CV+C+CC+VV~RCI}	          &1627.25& 9.69183E-03& 9.28846E-03& 7.98E-03& 4.28E-03& -6.58296E+01& B+ \\
\textbf{RCI~(accum) 0.99996}   &1626.84& 9.43809E-03& 8.82439E-03& 7.91E-03& 4.19E-03& -4.13651E+01& B+ \\
\textbf{RCI~(accum) 0.99982}   &1625.28& 8.68898E-03& 7.58237E-03& 7.73E-03& 3.94E-03& -2.00631E+01& C+ \\
\textbf{RCI~(accum) 0.9997}   &1624.30& 8.17001E-03& 6.80147E-03& 7.61E-03& 3.79E-03& -1.47317E+01& C \\
\textbf{RCI~(accum) 0.9995}   &1622.91& 7.44070E-03& 5.82335E-03& 7.36E-03& 3.55E-03& -1.08477E+01& D+ \\
\textbf{RCI~(accum) 0.99947}   &1622.66& 7.33234E-03& 5.67099E-03& 7.32E-03& 3.51E-03& -1.03166E+01& D+ \\
\textbf{RCI~(RSMBPT) 100\%}       &1627.33& 9.69344E-03& 9.28389E-03& 7.98E-03& 4.28E-03& -6.48160E+01& B+ \\
\textbf{RCI~(RSMBPT) 99.95\%}	    &1627.02& 9.68893E-03& 9.19334E-03& 7.99E-03& 4.26E-03& -5.31666E+01& B+ \\
\textbf{RCI~(RSMBPT) 99.5\%}      &1623.63& 9.70334E-03& 8.94842E-03& 7.96E-03& 4.16E-03& -3.42194E+01& B  \\
\textbf{RCI~(RSMBPT) 99\%}	      &1620.14& 9.69110E-03& 8.92997E-03& 7.90E-03& 4.15E-03& -3.38773E+01& B  \\
\textbf{RCI~(RSMBPT) 95\%}	      &1599.27& 9.83055E-03& 7.95396E-03& 7.78E-03& 3.75E-03& -1.26580E+01& C  \\
\multicolumn{8}{c}{$\mathrm{4s^24p5p~^1S_0}$ -- $\mathrm{4s^24p5d~^1P^o_1}$} \\
\textbf{CV+C+CC+VV~RCI}	          &2056.26& 1.28936E+01& 6.02123E+00& 4.82E-01& 3.37E-01& -3.05221E+00& E \\
\textbf{RCI~(accum) 0.99996}   &2050.73& 1.28697E+01& 5.98194E+00& 4.83E-01& 3.38E-01& -3.02977E+00& E \\
\textbf{RCI~(accum) 0.99982}   &2035.90& 1.28063E+01& 5.87452E+00& 4.84E-01& 3.41E-01& -2.96810E+00& E \\
\textbf{RCI~(accum) 0.9997}   &2025.87& 1.27665E+01& 5.80378E+00& 4.84E-01& 3.43E-01& -2.92718E+00& E \\
\textbf{RCI~(accum) 0.9995}   &2011.53& 1.27078E+01& 5.69493E+00& 4.85E-01& 3.45E-01& -2.86398E+00& E \\
\textbf{RCI~(accum) 0.99947}   &2009.55& 1.26997E+01& 5.68012E+00& 4.85E-01& 3.45E-01& -2.85547E+00& E \\
\textbf{RCI~(RSMBPT) 100\%}       &2056.18& 1.28927E+01& 6.02228E+00& 4.82E-01& 3.37E-01& -3.05342E+00& E \\
\textbf{RCI~(RSMBPT) 99.95\%}	    &2056.33& 1.28920E+01& 6.02323E+00& 4.82E-01& 3.38E-01& -3.05445E+00& E \\
\textbf{RCI~(RSMBPT) 99.5\%}      &2058.84& 1.28848E+01& 6.02658E+00& 4.82E-01& 3.39E-01& -3.05982E+00& E \\
\textbf{RCI~(RSMBPT) 99\%}	      &2062.39& 1.28808E+01& 6.04391E+00& 4.83E-01& 3.40E-01& -3.07530E+00& E \\
\textbf{RCI~(RSMBPT) 95\%}	      &2094.38& 1.28651E+01& 6.19306E+00& 4.82E-01& 3.43E-01& -3.20467E+00& E \\
\multicolumn{8}{c}{$\mathrm{4s^24p^2~^3P_0}$ -- $\mathrm{4s4p^3~^3D^o_1}$} \\
\textbf{CV+C+CC+VV~RCI}	          &914.84& 1.34217E-01& 1.02639E-01& 1.74E-02& 8.49E-03& -9.85309E+00& D+ \\
\textbf{RCI~(accum) 0.99996}   &913.83& 1.34955E-01& 1.02857E-01& 1.75E-02& 8.53E-03& -9.72275E+00& D+ \\
\textbf{RCI~(accum) 0.99982}   &910.29& 1.36953E-01& 1.03277E-01& 1.78E-02& 8.63E-03& -9.33166E+00& D+ \\
\textbf{RCI~(accum) 0.9997}   &907.44& 1.38392E-01& 1.03355E-01& 1.79E-02& 8.69E-03& -8.99907E+00& D+ \\
\textbf{RCI~(accum) 0.9995}   &903.05& 1.40476E-01& 1.03508E-01& 1.82E-02& 8.80E-03& -8.57272E+00& D+ \\
\textbf{RCI~(accum) 0.99947}   &902.38& 1.40770D-01& 1.03504D-01& 1.82E-02& 8.81E-03& -8.50874E+00& D+ \\
\textbf{RCI~(RSMBPT) 100\%}       &914.83& 1.34206E-01& 1.02620E-01& 1.74E-02& 8.48E-03& -9.84897E+00& D+ \\
\textbf{RCI~(RSMBPT) 99.95\%}	    &914.91& 1.34091E-01& 1.02576E-01& 1.74E-02& 8.49E-03& -9.86616E+00& D+ \\
\textbf{RCI~(RSMBPT) 99.5\%}      &915.65& 1.33498E-01& 1.02287E-01& 1.74E-02& 8.47E-03& -9.92974E+00& D+ \\
\textbf{RCI~(RSMBPT) 99\%}	      &916.35& 1.33157E-01& 1.02251E-01& 1.74E-02& 8.47E-03& -1.00183E+01& D+ \\
\textbf{RCI~(RSMBPT) 95\%}	      &922.38& 1.32410E-01& 1.05113E-01& 1.74E-02& 8.73E-03& -1.15580E+01& D+ \\
\multicolumn{8}{c}{$\mathrm{4s^24p^2~^3P_0}$ -- $\mathrm{4s^24p4d~^3D^o_1}$} \\
\textbf{CV+C+CC+VV~RCI}	          &627.84& 6.14607E+00& 4.71096E+00& 4.17E-01& 3.94E-01& -9.94495E+00& D+ \\
\textbf{RCI~(accum) 0.99996}   &627.45& 6.14926E+00& 4.70313E+00& 4.18E-01& 3.94E-01& -9.85846E+00& D+ \\
\textbf{RCI~(accum) 0.99982}   &626.04& 6.15669E+00& 4.67532E+00& 4.21E-01& 3.97E-01& -9.58524E+00& D+ \\
\textbf{RCI~(accum) 0.9997}   &624.87& 6.15789E+00& 4.64922E+00& 4.24E-01& 3.98E-01& -9.37375E+00& D+ \\
\textbf{RCI~(accum) 0.9995}   &623.02& 6.15485E+00& 4.60592E+00& 4.27E-01& 4.01E-01& -9.06659E+00& D+ \\
\textbf{RCI~(accum) 0.99947}   &622.75& 6.15570E+00& 4.60079E+00& 4.28E-01& 4.01E-01& -9.02472E+00& D+ \\
\textbf{RCI~(RSMBPT) 100\%}       &627.83& 6.14447E+00& 4.70959E+00& 4.17E-01& 3.93E-01& -9.94365E+00& D+ \\
\textbf{RCI~(RSMBPT) 99.95\%}	    &627.85& 6.14266E+00& 4.70825E+00& 4.17E-01& 3.93E-01& -9.94412E+00& D+ \\
\textbf{RCI~(RSMBPT) 99.5\%}      &628.25& 6.14983E+00& 4.72028E+00& 4.18E-01& 3.94E-01& -9.99964E+00& D+ \\
\textbf{RCI~(RSMBPT) 99\%}	      &628.65& 6.15474E+00& 4.72978E+00& 4.18E-01& 3.95E-01& -1.00488E+01& D+ \\
\textbf{RCI~(RSMBPT) 95\%}	      &631.76& 6.15534E+00& 4.78122E+00& 4.19E-01& 4.01E-01& -1.05039E+01& D+ \\
\end{longtable}
}

\section{Conclusions}

The method, based on the Rayleigh-Schr\"odinger perturbation theory in an irreducible tensorial form, 
is extended to estimate the VV correlations. The expressions to calculate the influence of these correlations are provided.
This extended RSMBPT method allows to estimate the contribution 
of any $K'$ configuration of the CV, C, CC and VV correlations 
with preferred core, virtual orbitals sets and with any number of valence electrons for any atom or ion.

The developed RCI (RSMBPT) method works perfectly, when the CV, C, CC and VV correlations 
are accounted for using the RSMBPT method, the results reproduce the regular {\sc Grasp}2018 data.
Also, this method works very well even when the MR set consists 
of orbitals with few different principal quantum numbers, namely with $n$, $n+1$, and $n+2$ (where $n$=4).
The main advantages of the developed method in comparison to the regular method are as follows:  it allows to identify 
the most important CV, C, CC and VV correlations and significantly reduces the CSF space. 
These benefits of the RSMBPT method over the regular method would be helpful and valuable 
for calculations involving complex atoms and ions.

By employing the RSMBPT method to analyse only the results of the smaller bases for example
with 95\% and 99\% correlations, it is possible to extrapolate very accurately the total value of the energy 
that would be obtained by including 100\% of the correlations.
The extrapolation of the total energies would be helpful for complex computations 
when considering CV, C, CC and VV correlations, as such computation lead to large CSFs basis and are time consuming.
This could be very important for the study of energy levels that are close to each other.

Comparing the \textbf{CV+C+CC+VV~RCI~(RSMBPT)} 
and \textbf{CV+C+CC+VV~RCI~(accum)} results with the regular \textbf{CV+C+CC+VV~RCI} results, 
the differences between the results using RSMBPT and the regular {\sc Grasp}2018 energies 
are similar for all computed levels while the differences between the results using 'rmixaccumulate' program 
and the regular ones are scattering for all levels.
The reduction of CSFs basis in both computations is different, 
the selection process in RSMBPT is based on configurations meanwhile ’rmixaccumulate’
program is based on CSFs.

Most theoretical methods have a CI or RCI stage. It is usually at this stage that the size of the CSF base in terms of computing capacity is decided. Due to the complexity of the computation, a usually restricted active space (RAS) approach is used, which is then further reduced on the basis of one or the other theoretical principles. Regular GRASP2018 calculations also use the RAS approach and it will be reduced later using: 'rmixaccumulate', 'rcsfinteract', and 'cndens' programs \cite{Jonatal:2023b,grasp2007}. How these or other RAS reduction mechanisms work for TE, and what kind of error can be expected, can only be known after a full study. Meanwhile, using the RSMBPT correlation selection method to reduce the RAS space, we also determine the resulting TE error (difference of TE computed in full RAS and reduced RAS) created by the RAS space reduction. As we know, this theoretical method is the only one that allows us to predict the errors of reduction of the RAS to the TE on the basis of the calculation parameters.

\bibliographystyle{pccp}

\end{document}